\documentclass[11pt,letterpaper]{article}
\pdfoutput=1
\usepackage{jcappub}
\usepackage{subcaption}
\usepackage{bbm,bm}
\usepackage{mathrsfs}
\usepackage{slashed}
\usepackage{caption}
\usepackage{epstopdf}
\usepackage[normalem]{ulem}
\usepackage[bottom]{footmisc}
\usepackage{subcaption}
\usepackage{bbold}
\usepackage{titlesec}
\usepackage{threeparttable}
\usepackage{booktabs}
\usepackage{changepage}
\usepackage[utf8]{inputenc}
\usepackage{soul}
\usepackage[greek.polutoniko,english]{babel}

\usepackage{grffile}

\usepackage{graphicx}  % needed for figures
\usepackage{dcolumn}   % needed for some tables
\usepackage{bm}        % for math
\usepackage{amssymb}   % for math
\usepackage{setspace}
\usepackage{amsmath, amssymb, setspace}
\usepackage{array}
\usepackage{booktabs}
\usepackage{caption}
\usepackage{indentfirst}
\usepackage{float}
\usepackage{lmodern}
\usepackage{multirow}
\usepackage{soul}
\usepackage[normalem]{ulem}

\usepackage{braket}
\usepackage{comment}

\usepackage[draft]{pgf}

\newcommand{\DoBox}[1]{\begin{center}
\color{red}\fbox{
\begin{minipage}{0.9\textwidth}

\end{minipage}}
\end{center}}

\newlength{\myimageoversize}
\newsavebox{\myimage}

\titleformat{\subsubsection}
 {\normalfont\fontsize{12}{17}\itshape}{\thesubsubsection}{1em}{}

\title{\huge{Catastrogenesis: DM, GWs, and PBHs from ALP string-wall networks}}

\author[a]{Graciela B. Gelmini,}
\author[a]{Anna Simpson,}
\author[a]{Edoardo Vitagliano}

\affiliation[a]{Department of Physics and Astronomy, University of California, Los Angeles\\
475 Portola Plaza, Los Angeles, CA 90095-1547, USA}

\emailAdd{gelmini@physics.ucla.edu}
\emailAdd{ansimps@g.ucla.edu}
\emailAdd{edoardo@physics.ucla.edu}

\begin{document}

%%%%%%%%%%%%%%%%%%%%%%%%%%%%%%%%%%%
\abstract{
Axion-like particles (ALPs), a compelling candidate for dark matter (DM), are the pseudo Nambu-Goldstone bosons of a spontaneously and explicitly broken global $U(1)$ symmetry. When the symmetry breaking happens after inflation, the ALP cosmology predicts the formation of a string-wall network which must annihilate early enough, producing gravitational waves (GWs) and primordial black holes (PBHs), as well as non-relativistic ALPs. We call this process \textit{catastrogenesis}.
We show that, under the generic assumption that the potential has several degenerate minima, GWs from string-wall annihilation at temperatures below 100 eV could be detected by future CMB and astrometry probes, for ALPs with mass from $10^{-16}$ to $10^{6}$~eV. In this case, structure formation could limit ALPs to constitute a fraction of the  DM and the annihilation would produce mostly ``stupendously large" PBHs.   For larger annihilation temperatures, ALPs can constitute $100\%$ of DM, and the annihilation could produce  supermassive black holes with a mass of up to $10^9\, M_\odot$ as found at the center of large galaxies. Therefore our model can solve two mysteries, the nature of the DM and the origin of these black holes.
}
%%%%%%%%%%%%%%%%%%%%%%%%%%%%%%%%%%% 
%%%%%%%%%%%%%%%%%%%%%%%%%%%%%%%%%%%

\maketitle

\section{Introduction }

\begin{figure}[t]\centering\vspace{-5em}\hspace{-1em}
\includegraphics[width=0.78\textwidth]{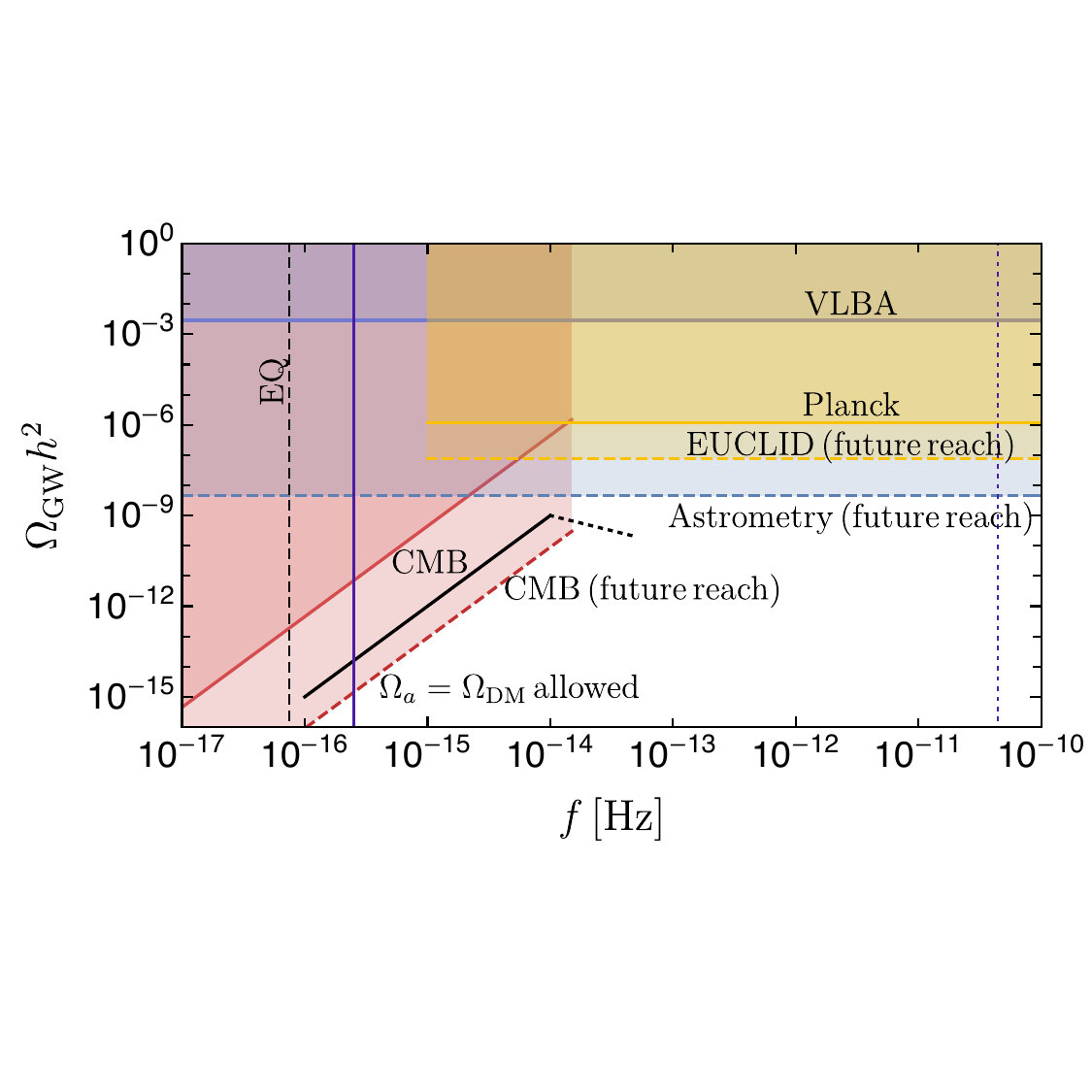}
\vspace{-4em}
\caption{Regions in $\Omega_{\rm GW}h^2$ as a function of GW frequency, excluded by existing bounds (solid colored lines) or within the expected reach  (dashed colored lines) of $N_{\mathrm{eff}}$~\cite{Pagano:2015hma,Laureijs:2011gra} (yellow), astrometry~\cite{Darling:2018hmc,Arvanitaki:2019rax} (blue) and CMB~\cite{Namikawa:2019tax} (red) measurements. An example of the predicted peaked differential spectrum from string-wall annihilation with $T_{\rm ann}\simeq 100$ eV and $\Omega_{\rm GW}h^2|_{\rm peak}\simeq 10^{-9}$ (see main text) is shown in black: 
the $\sim f^3$ spectrum (solid black line) below the peak is predicted by causality and the $\sim f^{-1}$ spectrum above the peak (dotted black line) is uncertain.  The vertical dashed line labeled ``EQ" indicates the frequency of  GWs produced at matter-radiation equality.
If the peak is located to the left of the vertical dark blue solid line, ALPs cannot constitute  all of the DM due to CMB bounds alone. Structure formation bounds may shift the allowed region for ALPs constituting 100\% of the DM to much larger frequencies, to the right of the vertical dotted blue line. However, $\Omega_a~\lesssim~ 0.3~\Omega_{\rm DM}$ is allowed for the entirety of the parameter space shown.
}\label{fig:differentialspectrum}
\end{figure}

Gravitational waves (GWs) constitute a new window into our Universe, and simultaneously a new  tool to test particle physics models~\cite{Vilenkin:2000jqa,Maggiore:1900zz,Maggiore:2018sht,Sathyaprakash:2009xs,Barack:2018yly}. Recently we showed~\cite{Gelmini:2021yzu} that a particular class of light bosonic dark matter (DM) particle candidates could be tested by cosmological measurements in the near future of a stochastic GW background in the frequency range $10^{-16}- 10^{-14}~\mathrm{Hz}$, characterized by a peaked spectrum.
As shown in Fig.~\ref{fig:differentialspectrum}, progress is expected in this frequency range from several types of measurements, in particular from cosmic microwave background (CMB) temperature and polarization observations, taken from Ref.~\cite{Namikawa:2019tax} (see the red region in Fig.~\ref{fig:differentialspectrum}). 
 
Here we study further the type of model of Ref.~\cite{Gelmini:2021yzu}, in particular recognizing the importance of structure formation limits, and point out that they could also produce primordial black holes (PBHs).
 
The existence of a spontaneously broken global $U(1)$ symmetry is postulated  in many  extensions of the Standard Model (SM) of elementary particles, such as in
the original axion model~\cite{Peccei:1977hh, Weinberg:1977ma, Wilczek:1977pj}, invisible axion (also called QCD~axion) models~\cite{Kim:1979if,Shifman:1979if,Dine:1981rt,Zhitnitsky:1980tq},  Majoron models~\cite{Chikashige:1980ui,Gelmini:1980re},  familon models~\cite{Wilczek:1982rv,Reiss:1982sq,Gelmini:1982zz}  and axion-like particle (ALP) models (e.g.~\cite{Svrcek:2006yi,Arvanitaki:2009fg,Acharya:2010zx,Dine:2010cr,Jaeckel:2010ni}). The spontaneous breaking of a global symmetry produces a Nambu-Goldstone (NG) boson for every broken generator of the symmetry---one in the case of a $U(1)$.  Typically, in the models just mentioned, the $U(1)$ global symmetry is explicitly broken into a discrete subgroup at an energy scale $v$ much lower than that of the spontaneous breaking of the symmetry, $V.$ In this case, the NG boson, which would be massless without this breaking,  acquires a mass $m_a \simeq v^2/V$ and could constitute  part or all of the DM. We will call the pseudo-NG boson an axion-like particle (ALP)  and denote it with $a$. This particle could have very small couplings to the SM or be part of an entirely dark sector with no connection to the SM.

Already in the 1980s (see e.g. Ref.~\cite{Vilenkin:1984ib} and references therein) it was understood that if the spontaneous breaking of the global $U(1)$ symmetry happens after inflation, as we assume here, a system of cosmic walls bounded by strings appears due to the explicit symmetry breaking. First,  global cosmic strings are created at the spontaneous breaking, at a temperature scale $T \simeq V$. Later, once the lifetime of the Universe becomes comparable to the  oscillation period of the $a$ field about the closest minimum of its potential,  the global strings  become connected by walls. This happens at a time $t \simeq m_a^{-1}$. The characteristics of the string-wall system so produced depend on the number $N$ of minima of the explicit breaking potential.

If the explicit breaking potential has only one minimum, $N=1$, each string is at the border of just one wall which terminates in another string, forming separate wall ``ribbons" surrounded by true vacuum. These ribbons shrink due to the pull of the walls on the strings, and thus the string-wall system decays immediately after their formation.

If instead the explicit breaking potential has several degenerate minima, a number $N>1$ of them, each string connects to several walls, forming a stable string-wall system, which however must somehow be unstable due to cosmological constraints.

In their seminal paper dating back to 1974,  Zel'dovich, Kobzarev, and Okun~\cite{Zeldovich:1974uw} showed that the formation of a system of walls in the early Universe is  cosmologically unacceptable, unless the walls disappear early enough as to never dominate the energy density of the Universe. They proposed as a solution that a small energy difference between the minima of the potential at both sides of each wall, a ``bias", could drive the walls to annihilate early enough.  A generic  additional term in the explicit breaking scalar potential was thus later proposed~\cite{Sikivie:1982qv} as a way to implement this solution  for the  QCD axion string-wall system (see e.g. Ref.~\cite{Chang:1998bq}). We here adopt a similar term for ALPs. This extra term produces a small bias  among the $N$ minima leading to just one true vacuum. In due time this bias becomes dynamically important and accelerates each wall towards its adjacent higher energy vacuum, driving the evolution of the domain walls towards their annihilation (see e.g. Ref.~\cite{Gelmini:1988sf}). 
Here we will parameterize the bias as $V_{\rm bias} \simeq \epsilon_b v^4$ with a dimensionless positive parameter $\epsilon_b \ll 1$.

ALPs and GWs are produced by the cosmic strings before walls appear connecting them,  and if $N>1$ also by the string-wall system. GWs potentially produced solely by cosmic strings have been recently studied both for NG boson models without an explicit breaking, i.e. in which the bosons are massless~\cite{Chang:2019mza},  and for ALPs  assuming N=1~\cite{Gorghetto:2021fsn}, in which case the cosmic string system is terminated as soon as walls appear. The latter would be observable in future GW detectors if $V\gtrsim 10^{14}~\rm{GeV}$~\cite{Gorghetto:2021fsn}. For $N>1$, GWs could also be observable in GW detectors in multiple axion models (see e.g.~\cite{Higaki:2016jjh}) or if the axion is very heavy and decays fast enough (see e.g.~\cite{ZambujalFerreira:2021cte}).\footnote{If the walls are not bounded by strings, additional signatures are possible~\cite{Takahashi:2020tqv,Kitajima:2022jzz}.}

As in Ref.~\cite{Gelmini:2021yzu}, here we focus on models with $N>1$, in which case, for small enough values of the bias, i.e. of $\epsilon_b$, the dominant production of both GWs and stable, not extremely light ALPs happens when the string-wall system annihilates---a production we dub ``catastrogenesis''.\footnote{We name it after the Greek word \textgreek{katastrof\'h}, ``overturn'' or ``annihilation''.} Additionally, after the annihilation of the string-wall system starts,  some fraction of the closed walls could
shrink to their Schwarzschild radius and collapse into
PBHs~\cite{Ferrer:2019pbh}.

It is important to notice that the string-wall system annihilates after having entered into a ``scaling regime" in which the typical linear size of the walls is the horizon size, i.e. the lifetime of the Universe $\simeq t$, which erases the dependence of the system on how it was generated. 
Annihilation starts when the contribution of the volume energy density $V_{\rm bias}$ becomes comparable to, and later dominant over,  the wall energy density, characterized by  the wall surface tension $\sigma \sim v^2 V$, at which point the volume pressure accelerates the walls towards each other.   As a result the density of GWs, ALPs and PBHs produced at annihilation depend only on two parameters, $V_{\rm bias}$ and $\sigma$ (or others related to these two) instead of the
three characterizing the initial potential of the model, which can be chosen to be e.g. $V$, $v$ and $\epsilon_b$, or  $V$, $m_a$ and $\epsilon_b$.

We will show that, depending on the annihilation temperature of the string-wall network, there are different observational consequences. If the annihilation happens close to recombination, our scenario is constrained by structure formation, though we find that ALPs can constitute a fraction up to 0.3 of the DM and the annihilation produces up to ``stupendously large" PBHs~\cite{Carr:2020erq,Deng:2021edw} which need to be necessarily intergalactic. In this case the annihilation produces GWs of frequencies $10^{-16}$ to $10^{-14}$ Hz that could be detected by CMB probes or astrometry measurements. On the other hand, for larger annihilation temperatures, above a few keV, ALPs can constitute $100\%$ of DM. In this case, if the annihilation happens at temperatures above $500\,\rm keV$, it can produce seeds for supermassive black holes (SMBHs) found at the center of large galaxies. Therefore, in the latter case, our model could solve two mysteries (the nature of DM and the production mechanism of SMBHs) in an economical manner.

We start in Section~2 by introducing a generic potential for pseudo-GBs, explain the cosmological history that it implies and define important parameters of the model such as the temperatures $T_w$ and $T_{\rm ann}$ at which walls form and annihilate, respectively.  The relic density of GWs and ALPs produced in this model are then computed in Section~3 and Section~4, respectively. The results of these last two sections are combined in Section~5 to explore the observability of GWs for different contributions of ALPs to the DM as a function of the ALP mass $m_a$.  We study the formation of PBHs in Section~6.  In Section~7 we explore Planck scale operators as the possible origin of the bias term in the ALP model potential, and in Section~8 we draw our conclusions.

\section{ALP models and their cosmology}

\begin{figure}  
  \centering
  \includegraphics[width=1\linewidth]{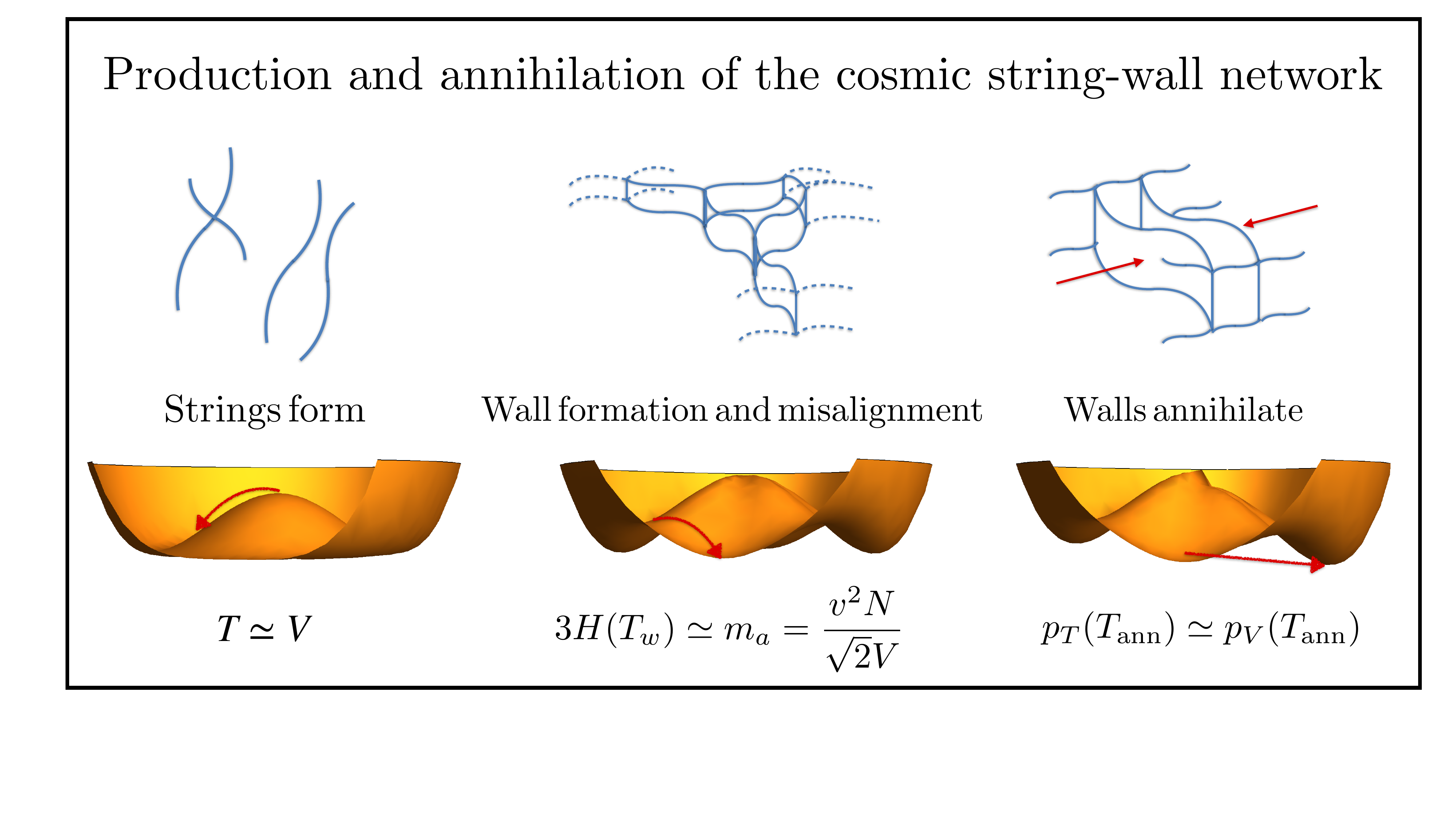}\vspace{-2em}
\caption{Illustration of the different stages of the ALP cosmology: spontaneous global $U(1)$ symmetry breaking at temperature $T\simeq V$ with the consequent production of cosmic strings; appearance of walls connected by the strings at the onset of the ALP field oscillations at temperature $T_w$  when the Hubble expansion is $H \simeq m_a/3$; and onset of the annihilation of the string-wall system at temperature $T_{\rm ann}$ when the volume pressure $p_V$ due to the potential bias becomes as important as, and later overcomes, the wall tension pressure $p_T$ and accelerates the walls towards the adjacent vacuum with higher energy.
}
\label{fig:ALP-cosmology}
\end{figure}

The generic potential $V(\phi)$ for the scalar field $\phi= |\phi | e^{i \theta}$ includes the terms, 
\begin{align}
\label{eq:potential}
V(\phi) \supset~ &  \frac{\lambda}{4} (|\phi |^2- V^2)^2 + \frac{v^4}{2} \left(1- \frac{|\phi |}{V} \cos(N \theta) \right) 
 -  \epsilon_b  v^4 \frac{|\phi |}{V}  \cos \left(\theta - \delta\right),
\end{align}
common to all models of the type we consider, with $V \gg v$.   The first term is $U(1)$ invariant and leads to the spontaneous breaking of this symmetry at a temperature scale $T \simeq V$, shortly after which the field vacuum expectation value is $|\phi |= V$ and the phase $\theta$ has different random values in different patches of the Universe. Cosmic strings are produced at this phase transition. Upper bounds on the energy scale of inflation~\cite{Hertzberg:2008wr,Aghanim:2018eyx} impose $V\lesssim 10^{16}~\mathrm{GeV}$. 

The relation $T \simeq V$ assumes that the bosons have the same temperature or average energy as visible sector particles. This could happen even if the field $\phi$ belongs to a dark sector not coupled to the SM, if the inflaton field responsible for inflation decays into dark particles with energy close to the inflaton mass, and the reheating temperature of the visible sector is also close to the inflaton mass, which would happen with a fast reheating.

  Shortly after the formation of cosmic global strings, the combined effect of the Hubble expansion and string recombination leads to a ``scaling" regime of the string system, in which the population of strings in a Hubble volume tends to remain of $\mathcal{O}(1)$ (see e.g. Ref.~\cite{Vilenkin:1984ib} and references therein).

The second and third terms in Eq.~\eqref{eq:potential} explicitly break the $U(1)$ symmetry of the first term.  The second term breaks the $U(1)$ symmetry into a $Z_{N}$ discrete subgroup, producing $N$ degenerate vacua along the previous $U(1)$ orbit of minima $|\phi|= V$. This gives a mass  
\begin{equation} \label{m-a}
m_a = \frac{ v^2 N}{\sqrt{2} V}
\end{equation}
to the ALP field $a= \theta V$ close to each minimum. We assume that  the couplings of the $a$ field are small enough that temperature corrections to $m_a$ are negligible. 

The equation of motion of the field $a$ in the expanding Universe is that of a harmonic oscillator with damping term $3H \dot{a}$, where $H= (2 t)^{-1}$ is the Hubble expansion rate during the radiation dominated epoch. 

While $3H \gg m_a$  the oscillation solution is overdamped and the field does not change with time. Only later, 
at a temperature $T_{w}$ when  $H(T_{w}) \simeq m_a/3$, regions of the Universe with different values of $\theta$ evolve to different minima. At this point, neighboring regions which happen to be in different vacua become separated by domain walls of  mass per unit surface (or surface tension)
\begin{equation} \label{sigma}
\sigma\simeq \frac{f_\sigma}{N} v^2 V.
\end{equation}
Here $f_\sigma$ is a model dependent dimensionless parameter. For $N=2$ the model is solvable analytically and one finds $f_\sigma \simeq 6$. Whenever choosing specific values of $N$ and $f_\sigma$ is required, i.e. to produce all our figures, we will assume $N=6$  and  $f_\sigma/N \simeq 1$.  

 The temperature  $T_{ w}$ when walls form is defined  by $H(T_w)\simeq m_a/3$, through the Friedmann equation giving the Hubble expansion rate as a function of the temperature for a radiation dominated universe,
\begin{equation} \label{def-T}
H(T) = \sqrt{\frac{\pi^2 g_\star(T)}{90}}~ \frac{T^2}{m_P},
\end{equation}
where $m_P= 2.4 \times 10^{18}$ GeV is the reduced Planck mass,  and $g_\star$ is the energy density number of degrees of freedom~\cite{Saikawa:2018rcs}.
This temperature  is 
\begin{equation} \label{Tw}
T_{ w} \simeq \frac{5 \times 10^4 \, {\rm GeV}}{\left[g_\star(T_w)\right]^{1/4}}  \left(\frac{m_a}{\rm eV}\right)^{1/2}.
\end{equation}
After a short time, the expansion of the Universe and energy losses drive the string-wall system into another ``scaling" regime in which the linear size of the walls is the cosmic horizon size $\simeq t$ . Therefore, its energy density at time $t$ is
\begin{equation} \label{rhow}
\rho_w \simeq \frac{\sigma}{t}.
\end{equation}

The third term in Eq.~\eqref{eq:potential}, proposed in Ref.~\cite{Sikivie:1982qv}, is assumed to be much smaller than the second one, i.e. $\epsilon_b \ll~1$. For the time being, we will remain agnostic about its origin.
It makes the vacuum closest to the arbitrary fixed phase $\delta$ the true vacuum, and rises the others by an energy density difference, a bias, of order  
\begin{equation} \label{Vbias}
V_{\rm bias} \simeq~\epsilon_b v^4.
\end{equation}

 The motion of the walls after they are formed is determined by the surface tension $\sigma$ and the energy difference $\simeq V_{\rm bias}$ between the vacua. The surface tension tends to rapidly straighten out curved walls to the horizon scale $H^{-1}$, and produces a pressure $p_T$.  It coincides with the energy density stored in the walls $p_T \simeq \rho_w$, which decreases with time. The volume pressure $p_V$, which coincides with the energy density difference between vacua $p_V \simeq V_{\rm bias}$,   tends to accelerate the walls towards their higher energy adjacent vacuum, converting the higher energy vacuum into the lower energy one. The energy released in this conversion fuels the wall motion. 
 
 Domain walls evolve differently depending on the relative size of $p_T$ and $p_V$ (see e.g. Ref.~\cite{Gelmini:1988sf}). Assuming  that  $p_V \ll  p_T$ when walls form (i.e. $\epsilon_b \ll 1$), when at a later time  $p_T$ approaches $p_V$, the bias drives the walls (and strings bounding them) to annihilate within a Hubble time.  
 
 Taking $p_T \simeq p_V$, i.e. $\rho_w \simeq \sigma/t_{\rm ann} \simeq V_{\rm bias} \simeq \epsilon_b v^4$, as the condition for wall annihilation, we obtain 
\begin{equation} \label{def-Tann}
H(T_{\rm ann})= \frac{1}{2 t_{\rm ann}} \simeq \frac{V_{\rm bias}}{2 \sigma} = \frac{\epsilon_b m_a}{\sqrt{2} f_\sigma},
\end{equation}
which defines the temperature $T_{\rm ann}$  at which the string-wall system annihilates, 
\begin{equation} \label{Tann}
T_{\rm ann} \simeq \frac{0.6 \times 10^5 \,{\rm GeV}}{[g_\star(T_{\rm ann})]^{1/4}}\, \sqrt{\frac{V_{\rm bias}}{\sigma \, {\rm eV}}}
\simeq \frac{0.7 \times 10^5 \,{\rm GeV}}{[g_\star(T_{\rm ann})]^{1/4}}\,  \sqrt{\frac{\epsilon_b \, m_a}{f_\sigma \, {\rm eV}}}.
\end{equation}
At this point the energy stored in the string-wall system is converted almost entirely into non-relativistic or mildly relativistic ALPs~\cite{Chang:1998tb}, but also into GWs and PBHs.

The different stages in the ALP cosmology just outlined are illustrated in Fig.~\ref{fig:ALP-cosmology}.

\section{Present GW energy density}

The quadrupole formula for the power emitted in GWs $P\simeq G\dddot{Q}_{ij}\dddot{Q}_{ij}$ is used to estimate the GW energy produced by the string-wall system~\cite{Maggiore:1900zz}.
 In the scaling regime, when the linear size of large walls is $\simeq~t$, the quadrupole moment of the walls as a function of the energy in the walls $E_{w} \simeq \sigma t^2$  is $Q_{ij} \simeq~E_{w} t^2$.  Thus $\dddot{Q}_{ij} \simeq \sigma t$, and the power emitted in GWs is  $P \simeq G \sigma^2 t^2$.  The energy density $\Delta \rho_{\rm GW}$ emitted in a time interval $\Delta t$ is then  
\begin{equation}
\label{Delta-rho-GW}
\Delta \rho_{\rm GW} (t) \simeq G \sigma^2 \frac{\Delta t} {t}. 
\end{equation}
Hence, in a time interval equal to the Hubble time $\Delta t \simeq t$, this energy density is always $G \sigma^2$,  independently of  the emission time $t$.
 The contribution of the waves emitted at the time $t$ to the present-day GW energy density is redshifted by the ratio $(R(t)/R_0)^4$, where $R(t)$ is the scale factor of the Universe at time $t$, and at present the scale factor is $R_0=1$. Therefore,  the largest contribution to the present GW energy density spectrum, namely the peak GW amplitude, corresponds to the latest emission time, the time $t= t_{\rm ann}$ at which the walls annihilate. Thus one has 
\begin{equation} \label{rho-peak-GW-walls}
\rho_{\rm GW}|_{\rm peak} \simeq G \sigma^2 \left(\frac{R(t_{\rm ann})}{R_0}\right)^4.
\end{equation}
 Defining as usual $\Omega_{\rm GW}h^2|_{\rm peak} = \rho_{\rm GW}|_{\rm peak} (h^2/\rho_c)$, where $\rho_c$ is the present critical density  and $h$ the reduced Hubble constant, using entropy conservation (i.e. that $g_{s \star} (t)(R(t) T(t))^3$ is constant) and $g_{s \star} (t_0)\simeq 3.93$~\cite{Saikawa:2018rcs} for the present the number of entropy degrees of freedom, we obtain

\begin{equation} \label{eq:OmegaGW-walls}
\Omega_{\rm GW}h^2|_{\rm peak} \simeq \epsilon_{gw}
\frac{1.2 \times 10^{-79} g_\star(T_{\rm ann})~ \sigma^4}{\left[g_{s \star}(T_{\rm ann}) \right]^{4/3} V_{\rm bias}^2{\text{GeV}}^4}
=\epsilon_{gw}
\frac{1.2 \times 10^{-79}  g_\star(T_{\rm ann})}{\epsilon_b^{2}~\left[g_{s \star}(T_{\rm ann}) \right]^{4/3}}
\left(\frac{f_\sigma V}{N {\text{GeV}}}\right)^4 .
\end{equation}
This estimate has been confirmed by numerical simulations~\cite{Hiramatsu:2010yz,Hiramatsu:2013qaa,Kawasaki:2011vv,Hiramatsu:2012sc,Hiramatsu:2013qaa}. Following Ref.~\cite{Hiramatsu:2012sc},  we include  in Eq.~\eqref{eq:OmegaGW-walls} a dimensionless factor $\epsilon_{gw} \simeq$  10 -- 20 for $N=6$ (see Fig.~8 of Ref.~\cite{Hiramatsu:2012sc}) found in numerical simulations that parameterizes the efficiency of GW production. For our figures we conservatively assume $\epsilon_{gw}=10$.   
 
 Notice that Eq.~\eqref{eq:OmegaGW-walls} defines also the maximum of the GW energy density spectrum at time $t$ as a function of the wave-number at present $k$ (which, when  defining $R_0=1$, coincides with the comoving wave-number) or of the frequency $f=k/(2\pi)$, which is defined as
\begin{equation}
\Omega_{\rm GW}h^2 (k, t) = \left(\frac{h^2}{\rho_c(t)}\right)
 \left(\frac{d \rho_{\rm GW} (t)}{d \ln k}\right) ,
\end{equation}
(see e.g. Refs.~\cite{Maggiore:1900zz,Gelmini:2020bqg}). Considering that in the scaling regime the characteristic frequency of the GWs emitted at $t$ is $\simeq H(t)$ (the inverse of the horizon size), the present-day frequency of waves emitted at time $t$ is $f \simeq R(t) H(t)$. For GWs emitted in the radiation dominated epoch, when $H(t)=(2t)^{-1}$,  $d \ln f= (H(t)- t^{-1}) dt$,  thus  $d \ln f = d \ln k =-(1/2)~ d \ln t$.  Using  from above that
$d \rho_{\rm GW} (t) \simeq  G \sigma^2 (d t/  t)$, we conclude that
\begin{equation}
\frac{d\rho_{\rm GW}(t)}{d\ln(k)}\simeq G \sigma^2,
\end{equation}
independently of $t$. Consequently, the peak amplitude of this GW spectrum at present, for $t= t_0$, coincides with the result in Eq.~\eqref{eq:OmegaGW-walls}.

Since the peak GW density is emitted at annihilation, its present frequency is $f_{\rm peak} \simeq R(t_{\rm ann}) H(t_{\rm ann})$, namely
\begin{equation} \label{f-peak}
f_{\rm peak}\simeq  0.76 \times 10^{-7} \text{Hz} ~\frac{T_{\rm ann}}{\rm GeV}~ \frac{\left[g_\star(T_{\rm ann})\right]^{1/2}}{\left[g_{s \star}(T_{\rm ann}) \right]^{1/3}}.
\end{equation}
The limit $T_{\rm ann} \gtrsim 5\,\rm eV$ (safely 
above matter-radiation equality, so that ALPs are produced early enough for them to be non-relativistic at matter-radiation equality) thus implies $f_{\rm peak} > 5\times 10^{-16}$ Hz.   As explained below, we find that the GWs observable in the near future in viable ALP models   should have $f_{\rm peak}$ close to this lower limit.

The order of magnitude estimate that we used to obtain the peak frequency in Eq.~\eqref{f-peak} is not sufficient to compute the spectrum of the GWs emitted by  cosmic walls. This spectrum has been computed numerically for $N>1$ in  Ref.~\cite{Hiramatsu:2012sc}. Figure~6 of Ref.~\cite{Hiramatsu:2012sc}   shows that the spectral slope changes at two scales: there is a peak at $k|_{\rm peak} \simeq R(t_f) m_a$ and there is a bump at the scale $k\simeq R(t_f) H(t_f)$ where $t_f$ is the latest time in their simulation.  Frequencies below the peak correspond to  super-horizon wavelengths at $t_{\rm{ann}}$, so causality requires the spectrum to go to zero as $k^3$ for $k < k_{\rm peak}$. Indeed, this is a characteristic of a white noise spectrum, as it corresponds to the absence of causal correlations~\cite{Caprini:2009fx}. The spectrum at frequencies above the peak depends instead on the particular model assumed to produce the GWs. The spectrum $1/k$ was found analytically for a source that is not correlated at different times, i.e. that consists of a series of short events~\cite{Caprini:2009fx}. The numerically obtained spectrum of Ref.~\cite{Hiramatsu:2012sc} has roughly a  $1/k$ dependence for $k > k_{\rm peak}$, although with approximate slope and height of the secondary bump that depend on $N$.

An example of the approximate spectrum just mentioned is shown in Fig.~\ref{fig:differentialspectrum}, together with several bounds and projected reaches of potential GW discovery in the near future. For $f> 10^{-14}$ Hz, the most important bounds are obtained with the Very Long Baseline Array (VLBA) astrometric catalog~\cite{Darling:2018hmc}, as GWs produce an apparent distortion of the position of background sources, and  from the effective number of neutrino species during CMB emission $N_{\mathrm{eff}}$~\cite{Pagano:2015hma}, as GWs are one of the radiation components in the early Universe. In the near future, EUCLID will improve bounds on $N_{\mathrm{eff}}$ by one order of magnitude~\cite{Laureijs:2011gra}, and astrometry could even reach $\Omega_{\mathrm{GW}}\simeq 10^{-8}$~\cite{Arvanitaki:2019rax}. At lower frequencies, measurements of the CMB polarization can be used to constrain GWs~\cite{Kamionkowski:1999qc,Smith:2005mm,Clarke:2020bil,Lasky:2015lej,Campeti:2020xwn}. Current bounds are obtained from Planck temperature~\cite{Aghanim:2018eyx} and BICEP/Keck Array polarization~\cite{Ade:2018gkx} data sets, and could be improved by planned experiments such as LiteBIRD~\cite{Matsumura:2013aja}, PICO~\cite{Hanany:2019lle}, and CORE~\cite{Delabrouille:2017rct}. Reference~\cite{Namikawa:2019tax}, whose constraints and projections we show, relaxed the usual assumption of a power-law background, and considered CMB constraints on monochromatic GWs, which may be closer to the peaked spectrum of our model. Notice that the constraints of Ref.~\cite{Namikawa:2019tax} are based on temperature anisotropies, while the projections for the future reach are obtained with the B-mode spectrum optimistically assuming a full sky observation, 1$\mu$K-arcmin white noise with 1 arcmin Gaussian beam.

  The dominant source of GW emission from the string network (before walls appear) are loops continuously formed by string fragmentation. Similarly to what is done for the string-wall system, the  estimates of the GW density emitted  are based on the quadrupole formula (see e.g. Refs.~\cite{Chang:2019mza,Gouttenoire:2019kij,Gorghetto:2021fsn} and references therein). The energy of the string network is $E_s \simeq \mu H^{-1}$, where $\mu$ is the string mass per unit length.  In this case, $\dddot{Q}_{ij} \simeq \mu$, and the power emitted in GWs is  $P \simeq G \mu^2$.  Using the same assumptions as for the walls, the energy density $\Delta \rho_{\rm GW}$ for strings is $\Delta \rho^{str}_{\rm GW} (t)\simeq G \mu^2 (\Delta t)$. 
  
  An approximate simple expression can fit the numerical spectra of GWs emitted
  by strings during the radiation dominated era obtained in Refs.~\cite{Chang:2019mza,Gouttenoire:2019kij,Gorghetto:2021fsn}, namely
\begin{equation} \label{OmegaGW-strings}
\Omega^{\rm st}_{\rm GW} h^2 \simeq 2 \times 10^{-15} \left(\frac{10^{-12} ~ {\rm Hz}}{f} \right)^{1/8} \left( \frac{V}{10^{14}~ {\rm GeV}} \right)^4.
\end{equation}
This spectrum has a low frequency cutoff at the frequency of GWs emitted by strings the latest, when the horizon size is largest. The Ly-$\alpha$ lower limit $m_a> 2 \times 10^{-20}$ eV~\cite{Rogers:2020ltq} on the mass of an ALP constituting all of the DM imposes a limit $T_w> 5.3$ keV (see Eq.~\eqref{Tw})  on the temperature at which walls appear and the system of only strings ceases to exist. The frequency at present of GWs emitted at $T_w$ by strings of typical size $H^{-1}$ is given by Eq.~\eqref{f-peak} using $T_w$ instead of $T_{\rm ann}$. This frequency is $4.7 \times 10^{-11}$ Hz for $T_w = 5.3$ keV,  and only larger frequencies are produced earlier. The spectrum cuts off at higher frequencies for larger $m_a$ (see Fig.~4 of Ref.~\cite{Gorghetto:2021fsn}). Therefore, in our model the only source of GWs with $f < 10^{-12}$ Hz is the string-wall system. 

\section{Present ALP energy density}

Applying to our model analytic derivations in the literature (see e.g. Refs.~\cite{Hiramatsu:2010yu, Hiramatsu:2012sc,Gouttenoire:2019kij,Gorghetto:2020qws} and references therein) we obtain the different components of the present ALP density. 

When the symmetry breaking happens after inflation, at the onset of the ALP field oscillations at temperature $T_w$,  when the Hubble expansion is $H \simeq m_a/3$,  different patches in the Universe have different initial phases $\theta_i= a_i/V$ misaligned with respect to the minima of the potential. Oscillations around these minima produce an ALP energy density at present (see e.g.~\cite{Preskill:1982cy,Abbott:1982af,Dine:1982ah})
\begin{equation} \label{Omega-a-missalig}
\Omega_a^{\rm mis} h^2=\left\langle\frac{1}{2}\theta_i^2 m_a^2 V^2 \left(\frac{R(T_w)}{R(t_0)}\right)^3 \frac{1}{\rho_c}h^2\right\rangle\simeq 2.5 \times 10^{-24} \langle \theta_i^2 \rangle ~  
\frac{V^2 m_a^{1/2}}{{\rm GeV}^{2} {\rm eV}^{1/2}}~
\frac{[g_\star(T_w)]^{3/4}}{g_{s \star}(T_w)}, 
\end{equation}
where $\langle \theta_{i}^2 \rangle=c_{\rm anh}\pi^2/3 $ is the (naive) average of $\theta_{i}$ over the present horizon volume, multiplied times the anharmonic coefficient $c_{\rm anh}\simeq 2$~\cite{Turner:1985si,Lyth:1991ub,Bae:2008ue,OHare:2021zrq}.

Soon after the spontaneous breaking of the $U(1)$ symmetry, the cosmic string system reaches its scaling regime and  emits ALPs continuously. The scaling solution is characterized by the parameter $\xi$, the average number of infinite strings in  a volume $t^3$, and thus the string energy density is $\rho_{\rm st} = \xi \mu/ t^2$. The string mass per unit length is (see e.g. Ref~\cite{Vilenkin:1984ib})
\begin{equation} \label{mu}
\mu \simeq 2 \pi V^2 \ln\left(\frac{t}{\sqrt{\xi}~ d_{\rm st}}\right).
\end{equation}
Here, the logarithm is due to the ALP long-range interactions between two strings, and includes the large distance cutoff given by the characteristic linear dimension $t/ \sqrt{\xi}$ of  strings and the small distance cutoff given by the string width, which we will approximate by $d_{\rm st} (t) \simeq 1/(\sqrt{2}~V)$~\cite{Hiramatsu:2010yu}. 

The computation of the present density of ALPs  produced by the cosmic strings reveals that it is dominated by the production at the time  $t_w$ when   walls appear and the emission by strings alone ceases. The walls appear when the effects of the axion mass become important, i.e. $t_w \simeq 3/(2 m_a)$, thus at earlier times the ALPs are effectively massless. The energy lost by strings goes dominantly into ALPs, so the number density $dn_e(t)$ of massless ALPs emitted by strings at time $t< t_w$ is given by 
\begin{equation} \label{dne}
dn^{\rm st}_e(t)  \simeq - \left(\frac{d\rho_{\rm st}(t)}{d t}\right) \frac{1}{\langle E_a \rangle},
\end{equation}
where ${\langle E_a \rangle}$ is the average ALP energy. The energy spectrum of axions from strings is sharply peaked at a momentum around inverse of the horizon scale and suppressed exponentially at higher momenta~\cite{Davis:1989nj, Yamaguchi:1998gx, Hiramatsu:2010yu}. Following Refs.~\cite{Yamaguchi:1998gx,Hiramatsu:2010yu} we approximate the average ALP energy by
${\langle E_a \rangle}\simeq 2\pi (4/t)$ (see e.g.  Appendix B and Eqs.~(27) and (42) of Ref.~\cite{Hiramatsu:2010yu}). Then,
taking into account the redshift to the present of $dn^{\rm st}_e(t)$ we can write the present number density of ALPs due to strings as 
\begin{equation} \label{na-strings}
n_a^{\rm st}(t_0)  \simeq - \int_{t_{\rm st}}^{t_w} dt \left(\frac{R(t)}{R_0}\right)^3  \left(\frac{d\rho_{\rm st}(t)}{d t}\right) \frac{t/4}{2 \pi}.
\end{equation}
Here $t_{\rm st}$ is the time at which strings appear, and performing the integral it is easy to see that for $t_{\rm st} \ll t_w$  it is dominated by its upper boundary, namely the production at $t_w$. 

The present energy density in non-relativistic ALPs produced by strings is $\rho_a^{\rm st}(t_0)= m_a n_a^{\rm st}(t_0)$, and taking $g_{s \star}(T_0)= 3.93$ and $g_{\star}(T_0)= 3.38$ as the present values of the energy and entropy density degrees of freedom we find
\begin{equation}
\label{Omega-a-strings}
\Omega_a^{\rm st} h^2 \simeq   \,  1 \,\times \, 10^{-23}\, \xi~
  \left(\frac{V}{\rm GeV}\right)^2 
\left(\frac{m_a}{\rm eV}\right)^{1/2}
\frac{[g_\star(T_w)]^{3/4}}{g_{s \star}(T_w)} \ln{\left(\frac{3V}{ \sqrt{2 \xi} ~ m_a}\right)}.
\end{equation}
Following Ref.~\cite{Gorghetto:2021fsn} we take $\xi= 25$ at the moment of wall annihilation (our Eq.~\eqref{Omega-a-strings} is not identical although very similar to the result in Eq. (19) of Ref~\cite{Gorghetto:2021fsn}). Therefore, the contribution to the ALP population from strings dominate over the population from misalignment mechanism alone (Eq.~\eqref{Omega-a-missalig}). Notice however that there are large uncertainties in the evaluation of the ALP population from cosmic strings~\cite{Gorghetto:2021fsn,OHare:2021zrq}, paralleling the discrepancy in the results of the QCD axion case~\cite{Kawasaki:2014sqa,Klaer:2017ond,Buschmann:2021sdq}.

When the bias parameter $\epsilon_b$ is not small enough, the contribution to the ALP density in Eq.~\eqref{Omega-a-strings} due to strings alone before walls appear  (plus misalignment),  can dominate over the contribution of the string-wall system that we compute below (see an example in the left panel of Fig.~\ref{fig:fixedmass6new}).  

The contribution to the ALP density due to the string-wall system is dominated by the emission at annihilation, which produces a number density of ALPs at present (see e.g. Section~4 of Ref.~\cite{Hiramatsu:2012sc})
\begin{equation} \label{n-a-walls}
n_a (t_0)\simeq  \left(\frac{R(T_{\rm ann})}{R_0}\right)^3 \frac{\rho_{ w}(T_{\rm ann})}{\langle E_a \rangle},
\end{equation}
where $\rho_w$ the string-wall system energy density, Eq.~\eqref{rhow}, and ${\langle E_a \rangle}$ is the average energy of the ALPs emitted by walls. Unlike those emitted by strings much earlier, these ALPs
are quasi non-relativistic,  $\langle p_a \rangle/m_a\simeq 1$, so that $\langle E_a\rangle \simeq \sqrt{2}m_a$~\cite{Hiramatsu:2012sc, Kawasaki:2014sqa}. As mentioned above, conservatively  we require  $T_{\rm ann}\gtrsim 5 \,\rm eV$, thus the ALP momentum at matter-radiation equality, i.e. when $T \simeq 0.75\,\rm eV$, is of order $\mathcal{O}(m_a/10)$ and ALPs are part of the CDM. The present energy density is 
 $\rho_a= n_a m_a$, and
\begin{equation} \label{Omega-a-walls}
\Omega_a h^2 \simeq  \frac{2 \times 10^{-42}{\rm eV^2}}{V_{\rm bias}^{1/2}}  \left(\frac{\sigma}{\rm eV^3}\right)^{3/2}
\frac{[g_\star(T_{\rm ann})]^{3/4}}{g_{s \star}(T_{\rm ann})}
\simeq \frac{2.4 \times 10^{-24} m_a^{1/2}}{\epsilon_b^{1/2} {\rm eV}^{1/2}}
\left(\frac{f_\sigma^{3/4} V }{N {\rm GeV}}\right)^2 
\frac{[g_\star(T_{\rm ann})]^{3/4}}{g_{s \star}(T_{\rm ann})} .
\end{equation}

Equations~\eqref{Omega-a-strings} and \eqref{Omega-a-walls} show that the string-wall contribution to the present ALP density dominates over that of the string system if $ \epsilon_b \lesssim 2 \times 10^{-9}$.

Combining Eqs.~\eqref{eq:OmegaGW-walls}, \eqref{f-peak} and  \eqref{Omega-a-walls}, the limit $\Omega_a h^2 < \Omega_{\rm DM} h^2 \simeq 0.12$
implies (neglecting degrees of freedom)
\begin{equation} \label{window}
\frac{\Omega_{\rm GW}h^2|_{\rm peak}}{10^{-15}} \left(\frac{f_{\rm peak}}{10^{-9} {\rm Hz}}\right)^2 < 10^{-4},
\end{equation}
which shows that our allowed observable window is at frequencies below the range of direct GW detection, which goes from $10^{-9}$ to $10^{3}$~Hz for $\Omega_{\rm GW}h^2 > 10^{-15}$. For example,  Eq.~\eqref{window} implies that for the reach of future astrometric data $\Omega_{\rm GW}h^2 \simeq 10^{-9}$, our observable window is at $f_{\rm peak} < 10^{-14}$~Hz. The differential spectrum with $\Omega_{\rm GW}h^2= 1 \times 10^{-9}$ and $T_{\rm ann}=100\,\rm eV$ that we show in Fig.~\ref{fig:differentialspectrum}, corresponding to $\sigma\simeq 200\,\rm GeV^3$ and $V_{\rm bias}\simeq 1 \times 10^{-30}\,\rm GeV^4$ (realized e.g. by $m_a\simeq 6\,\rm eV$, $\epsilon_b\simeq 4 \times 10^{-24}$, $V\simeq4\times 10^5\,\rm GeV$) saturates the bound given by Eq.~\eqref{window}.

\section{GW observability}

\begin{figure}
\begin{subfigure}{.5\textwidth}
  \centering
  \includegraphics[width=0.9\linewidth]{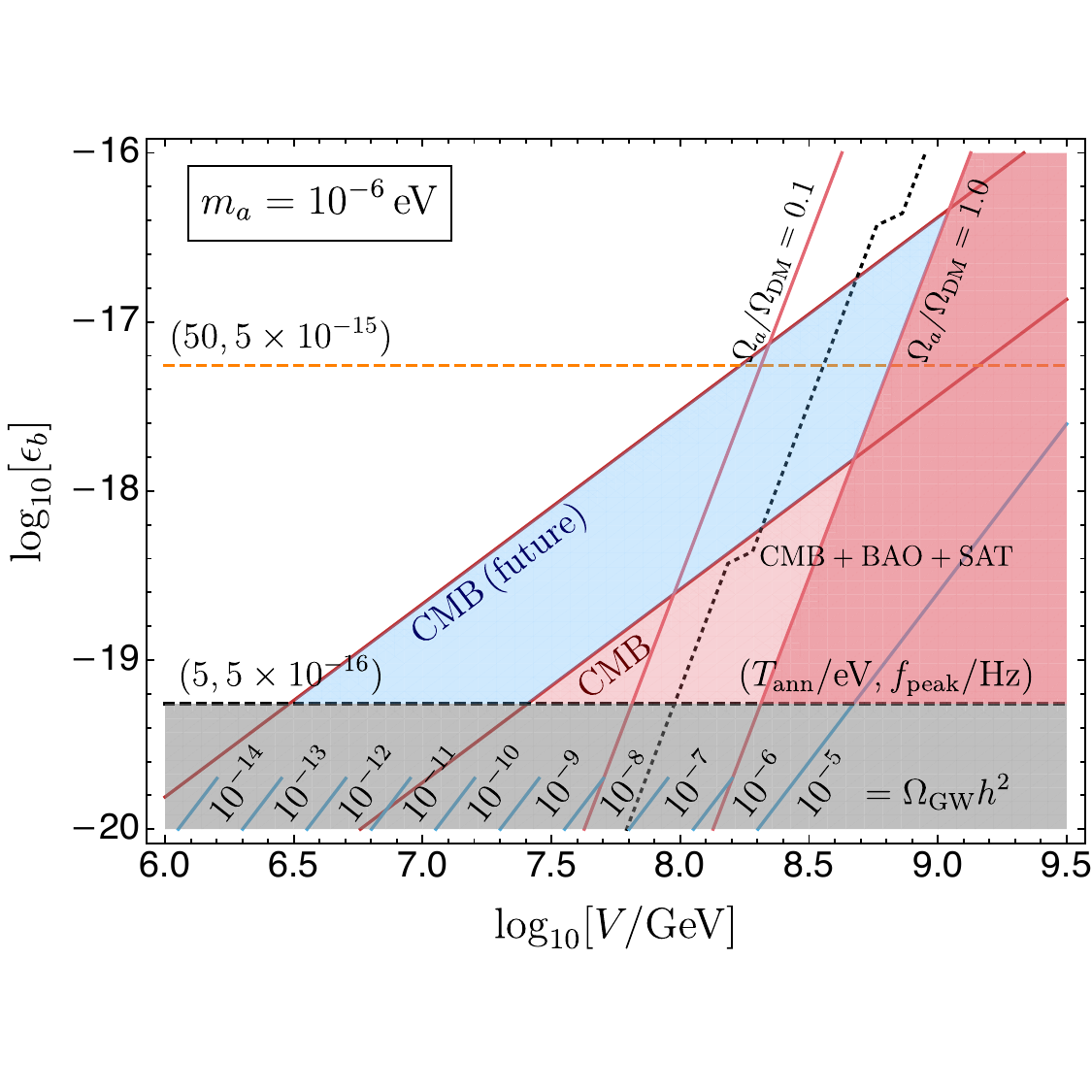}
  \label{fig:sfig1}
\end{subfigure}%
\begin{subfigure}{.5\textwidth}
  \centering
  \includegraphics[width=0.9\linewidth]{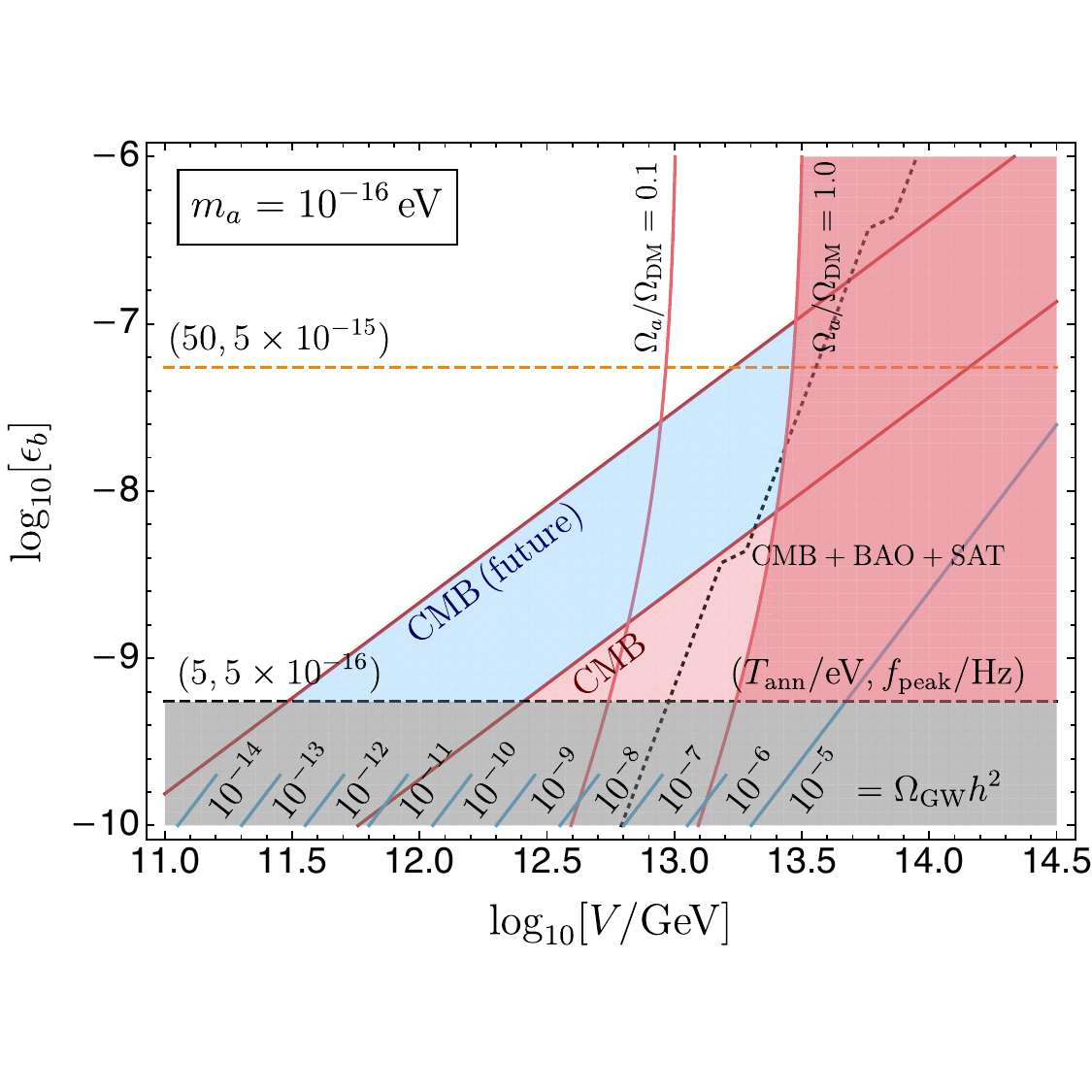}
  \label{fig:sfig2}
\end{subfigure}
\caption{Regions of interest of the bias parameter $\epsilon_b$ as a function  of  the  spontaneous  symmetry  breaking  scale $V$, for the ALP mass $m_a=10^{-6}$ eV (left panel) and  $m_a=10^{-16}$ eV (right panel). The red region is excluded either by an ALP density larger than that of DM, or by current CMB limits on GWs (see Fig.~\ref{fig:differentialspectrum}). Lines corresponding to   the fraction of the DM in ALPs $f_{\rm ALP} =\Omega_a/\Omega_{\rm DM}=1$ and $f_{\rm ALP}=0.1$
are also shown.  The grey region is excluded by requiring that ALPs be produced at $T_{\rm ann}>5$~eV. $T_{\rm ann}$ grows with $\epsilon_b$ as indicated by the $T_{\rm ann}=50$~eV orange dashed line.
The blue region will be explored in the near future by CMB probes and astrometry. We expect the region to the right of the black dotted lines to be subjected to structure formation bounds.
}
\label{fig:fixedmass6new}
\end{figure}

The region of the  $\{\epsilon_b,V\}$ parameter space allowed by all present bounds, and which can be explored by forthcoming measurements of low frequency GWs, depends on $m_a$. In Fig.~\ref{fig:fixedmass6new} we show two examples, for $m_a=10^{-6}\,\rm eV$ and $m_a=10^{-16}\,\rm eV$, respectively.  

The blue region in Fig.~\ref{fig:fixedmass6new} will be explored in the near future by CMB  and astrometry measurements. The GWs are observable at frequencies
\begin{equation} \label{f-window}
5 \times 10^{-16} {\rm Hz} <f_{\rm obs} <1 \times 10^{-14}  {\rm Hz},
\end{equation}
corresponding respectively to 
$5\,\rm eV <T_{\rm ann} < 10^2\,\rm eV$. Equations~\eqref{Tann} and \eqref{f-peak} show that $f_{\rm{peak}}$ is proportional to $\epsilon_b^2$, as shown in the figure. 

The red region in Fig.~\ref{fig:fixedmass6new} is excluded either by requiring the fraction of the DM in ALPs to be $f_{\rm ALP} =\Omega_a/\Omega_{\rm DM} \leq1$, or by current CMB limits on GWs (see Fig.~\ref{fig:differentialspectrum}).  The grey region corresponds to $T_{\rm ann} \lesssim 5\,\rm eV$. It is excluded by requiring that ALPs be produced early enough so  that all ALPs are non-relativistic at matter-radiation equality. 

It is easy to find how the observable region moves in the $\{\epsilon_b,V\}$ plane varying~$m_a$: it translates with $V \sim m_a^{-1/2}$ and  $\epsilon_b \sim m_a^{-1}$. 
  For a fixed ALP abundance, e.g. equal to the DM density,  Eq.~\eqref{Omega-a-walls} implies an expression for $V$ which can be substituted into Eq.~\eqref{eq:OmegaGW-walls}, to find that the GW amplitude depends on  $(m_a \epsilon_b)^{-1}\sim \sigma/V_{\rm bias}$. As shown in Eq.~\eqref{Tann} the annihilation temperature and thus the peak GW frequency in Eq.~\eqref{f-peak} depend on the ratio $V_{\rm bias}/\sigma \sim m_a \epsilon_b $ as well. 
  
  As $m_a$ increases, the lowest $V$ value of the observable region in Fig.~\ref{fig:fixedmass6new}, let us call  it $V_{\rm obs}$, decreases as
\begin{equation} \label{V-scaling-window}
 V_{\rm obs} \simeq  10^{6.5} {\rm GeV} \left(\frac{10^{-6} {\rm eV}}{m_a}\right)^{1/2}.
\end{equation}
  For consistency with the hierarchy of the terms in Eq.~\eqref{eq:potential} we require $v< 10^{-2} V$ and thus $m_a < 10^{-4} N V$. For $N=6$, compatibility with this limit restricts the observable window to  $V_{\rm obs}\gtrsim 2.5$ GeV and $m_a \lesssim 1.5$~MeV.
  
 The scaling of the characteristic bias of the observable region
\begin{equation} \label{eps-scaling-window}
\epsilon_{b,{\rm obs}}= 10^{-18} \left(\frac{10^{-6} {\rm eV}}{ m_a}\right),  
\end{equation}
 shows that ALP production by walls dominates over the production by strings  for 
 \begin{equation} \label{maWallsDominate}
m_a \gtrsim 5 \times 10^{-16}~{\rm eV},
\end{equation}
 for which $ \epsilon_b \lesssim 2 \times 10^{-9}$.
 Thus, the observable region in Fig.~\ref{fig:fixedmass6new} just translates with the same shape for $m_a\gtrsim 10^{-16}$ eV, until the contribution to the ALP population from strings becomes comparable to the wall contribution (as shown in the right panel).
 
 We show in Fig.~\ref{fig:fixedmass6new} the region to the right of the dotted black line  where we expect bounds coming from structure formation, specifically coming from CMB and baryon acoustic oscillation (BAO) measurements, and constraints on the number of Milky Way satellites.  Bounds on the late production of DM can be roughly estimated by using bounds on warm DM (WDM). The fraction of WDM to  total DM densities allowed by observations depends on the WDM mass $m_{\rm WDM}$: the smaller the mass, the lower the temperature  at which WDM particles become non-relativistic, $T\simeq m_{\rm WDM}/3$, and so become part of the cold DM. Similarly, the late production of cold DM in our scenario possibly implies large effects on CMB, BAO, and Milky Way satellite observations (for alternative realizations of late forming DM see e.g. \cite{Das:2006ht,Sarkar:2014bca,Das:2020nwc}).
 By taking $m_{\rm WDM}= 3\, T_{\rm ann}$ in Fig.~5 of Ref.~\cite{Diamanti:2017xfo}, we obtain the region of parameter space shown in Fig.~\ref{fig:fixedmass6new}, where structure formation bounds can be expected. A specific analysis of the structure formation in our scenario would be required to obtain actual limits. 

In the right panel of Fig.~\ref{fig:fixedmass6new}, the ALP mass is $m_a = 1 \times 10^{-16}$ eV and for this mass, as indicated in  Eq.~\eqref{maWallsDominate}, ALPs are  mostly produced through wall annihilation for $\epsilon_b\lesssim 2\times 10^{-9}$. However, we can see in the change of slope of the fixed $f_{\rm ALP}$ lines the dominance of the ALP population emitted by strings for larger $\epsilon_b$.
  
  Structure formation bounds are less stringent for ligher ALPs,  with $m_a \lesssim 10^{-16}\,\rm eV$,  dominantly
  produced by strings, since they are produced earlier, mostly at wall formation. For these ALPs $T_w > 0.5$ MeV $\gg T_{\rm ann}$. 
However, these lighter ALPs are subject to black hole superradiance limits (see e.g.~\cite{Mehta:2020kwu,Unal:2020jiy}).
These, together with structure formation bounds, are the only limits valid no matter how feeble the couplings of ALPs to SM  particles are, unless the ALP quartic self-coupling is large enough to suppress superradiance (see e.g. Ref.~\cite{Baryakhtar:2020gao}), in which case one can extend the observable window in Fig.~\ref{fig:fixedmass6new}  to lower ALP masses. Besides the bounds on $m_a\lesssim 10^{-16}\,\rm eV$ just mentioned, black hole superradiance also constrains the additional range  $10^{ - 13}\,\mathrm{eV}\lesssim m_a\lesssim 10^{-11}\,\mathrm{eV}$~\cite{Baryakhtar:2020gao}. 

The emission of GWs from the string-wall network does not depend on any coupling, be it ALP-photon, ALP-electron, ALP-nucleons, or CP-violating, see e.g. Refs.~\cite{Sikivie:2020zpn,Irastorza:2018dyq,OHare:2020wah}. Consequently, GWs could probe very ``dark" ALPs, which constitute a portal to a dark sector~\cite{Kaneta:2016wvf,Kalashev:2018bra,Arias:2020tzl,Caputo:2022npg}. Rather, the emission depends only on $\epsilon_b$ (for the GWs to be observable) and $m_a$. If future laboratory searches detect a particle through any coupling and find it to have a mass compatible with the QCD axion, the detection of GWs with a  spectrum similar to the one we described would challenge the attribution of this signal to a QCD axion, since GWs from QCD axions are not detectable~\cite{Hiramatsu:2012sc}.

\section{Primordial black holes} 

\begin{figure}  
  \centering
  \includegraphics[width=0.7\linewidth]{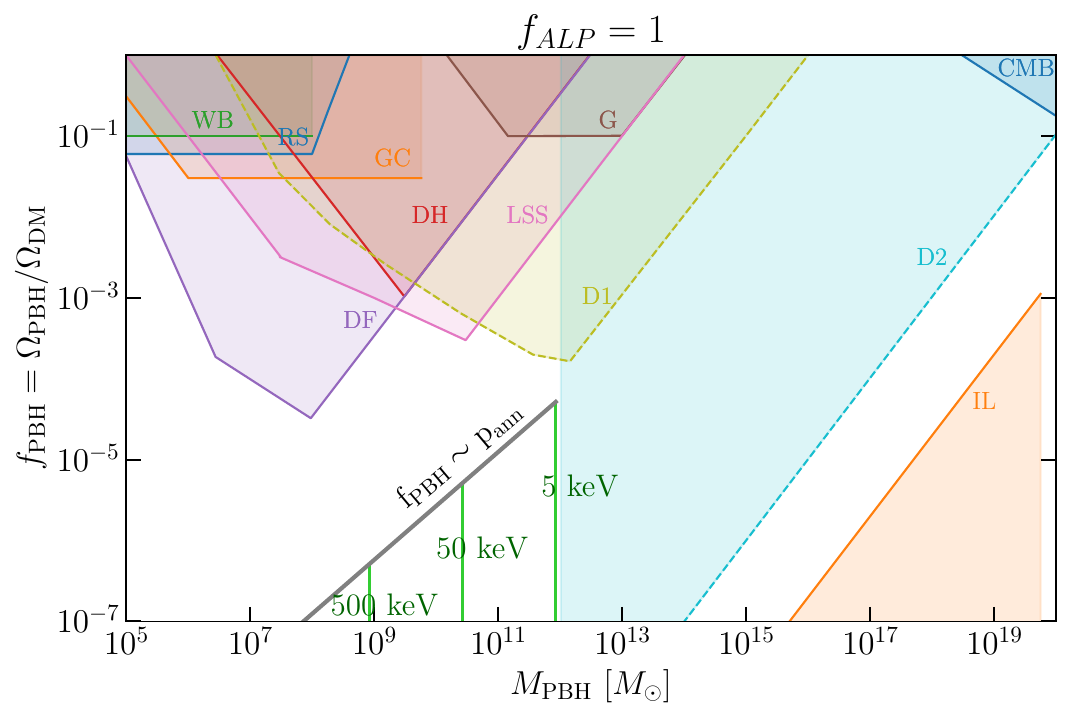}
  \includegraphics[width=0.7\linewidth]{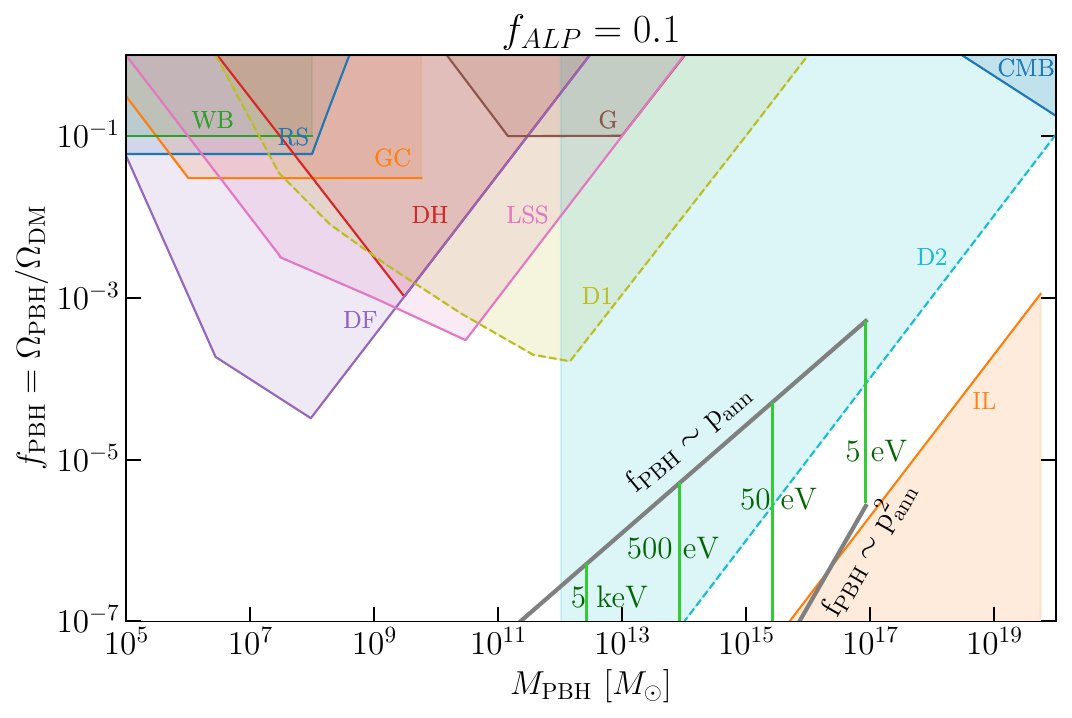}
\caption{Constraints on and predictions of our model for the density of PBHs, $f_{\rm PBH}=\Omega_{\rm PBH}/\Omega_{\rm DM}$, as a function of the PBH mass, assuming a monochromatic mass function. The excluded regions  (contoured by solid lines) taken from Ref.~\cite{Carr:2020pbh} and references therein, are due to: millilensing of compact radio
sources (RS), dynamical limits from disruption of wide binary stars (WB) and globular clusters (GC), heating of stars in the
Galactic disk (DH), dynamical
friction (DF), disruption of galaxies (G), and the CMB dipole (CMB). The incredulity limit
(IL) corresponds to one PBH per Hubble volume. Projected realistic (D1) and optimistic (D2) discovery limits through CMB spectrum $\mu$-distortion, taken from Ref.~\cite{Deng:2021edw}, are indicated by regions contoured by dashed lines. For a fixed ALP density ($f_{\rm ALP}=1$ in the upper panel and $f_{\rm ALP}=0.1$) in the lower panel,  our predicted PBH population could fall within the indicated funnel between solid gray lines, corresponding to the range in Eq.~\eqref{eq:fPBH}. The vertical green lines within each funnel show the PBH mass corresponding to each indicated value of the annihilation temperature $T_{\rm ann}$,  Eq.~\eqref{MPBH-Tann}. $f_{\rm ALP}=1$ is probably only possible with high enough annihilation temperatures, e.g. $T_{\rm ann}\gtrsim$ few keV, due to structure formation limits, while $f_{\rm ALP}=0.1$ is compatible with  $T_{\rm ann} \gtrsim\, \rm 5 eV$ (thus we cut the right portion of the funnels accordingly). 
}
\label{fig:PBH}
\end{figure}

During the process of annihilation of the string-wall system, some fraction of the closed walls could shrink to their Schwarzschild radius $R_{\rm Sch}(t)= 2 M(t)/M_P^2$ and collapse into primordial black holes (PBHs)~\cite{Ferrer:2019pbh}.  Here $M(t)$ is the mass within the closed wall at time $t$, and $M_P=1.22 \times 10^{19}$ GeV is the Planck mass. Considering that during the scaling regime the typical linear size of the walls is the horizon size $\simeq t$, PBH formation would happen if the ratio
\begin{equation}\label{eq:ptRadius}
    p(t)= \frac{R_{\rm Sch}(t)}{t}  =\frac{2 M(t)}{t\, M_P^2}
\end{equation}
is close to one, $p(t)\simeq 1$. As we will show, this could happen after the annihilation of the string-wall system has started, i.e. at $T\lesssim T_{\rm ann}$.

Annihilation starts when the contribution of the volume energy density to the mass within a closed wall of radius $t$  becomes as important as the contribution of the wall energy density. Shortly after, the volume density term dominates over the surface term, and the volume pressure accelerates the walls towards each other. 
 Close to annihilation, as a function of the lifetime of the Universe $t$, the mass within a closed wall is 
\begin{equation} \label{eq:M}
    M(t)\simeq \frac{4}{3}\pi t^3 V_{\rm bias} + 4\pi t^2 \sigma.
\end{equation}
Therefore, the ratio $p(t)$ increases with time. If at the moment of annihilation $p(t_{\rm ann})$ is  close to 1, PBHs would form as soon as the annihilation starts. However, we find that $p(t_{\rm ann}) \ll 1$, so PBHs could only form later, at a  time $t_* > t_{\rm ann}$ corresponding to a temperature $T_* < T_{\rm ann}$, for which $p(t_*)=1$. Temperature and time are related by $H=1/2t$ (see Eq.~\eqref{def-T}),  assuming radiation domination. At $T_*$, only a fraction of the original walls still remains. 

 The volume energy density due to the bias $V_{\rm bias} \simeq\epsilon_b v^4$ grows with time with respect to the  wall surface energy density, $\sigma/t$.   They become similar at $T_{\rm ann}$, $V_{\rm bias} \simeq\sigma/t_{\rm ann}$, therefore 
\begin{equation}
M(t_{\rm ann})\simeq \frac{16}{3} \pi t_{\rm ann}^3 V_{\rm bias},
\end{equation}
 so that Eq.~\eqref{eq:ptRadius} gives
\begin{equation}
   p(T_{\rm ann})\simeq 
   \frac{30}{\pi^2} \frac{V_{\rm bias}}{g_\star(T_{\rm ann})~T_{\rm ann}^4} .
\end{equation}
Notice that, as expected, $M(t_{\rm ann})$ and $p(T_{\rm ann})$ only depend on the parameters $V_{\rm bias}$ and $\sigma$.  As time elapses, for $T < T_{\rm ann}$, the volume contribution to the density dominates over the surface contribution, 
\begin{equation} \label{eq:M-late}
    M(t)\simeq \frac{4}{3}\pi t^3 V_{\rm bias} (1 + 3 \frac{t_{\rm ann}}{t}),
\end{equation}
 and 
\begin{equation} \label{eq:p-late}
 p(T)\simeq \frac{p(T_{\rm ann})}{4} 
 \left(\frac{t}{t_{\rm ann}}\right)^2 \left( 1 + 3 \frac{t_{\rm ann}}{t} \right).
\end{equation}
When $t\gg 3 ~t_{\rm ann}$ we can
 we neglect the second term in Eqs.~\eqref{eq:M-late} 
 and~\eqref{eq:p-late} and obtain 
\begin{equation}
 p(T)\simeq \frac{p(T_{\rm ann})}{4} \frac{g_\star(T_{\rm ann})}{g_\star(T)}  \left(\frac{T_{\rm ann}}{T}\right)^4.
\end{equation}
Using the last equation we find $T_*$, the temperature at which $p(T_*)=1$ in terms of $T_{\rm ann}$,
\begin{equation}
 p(T_*) \simeq \frac{p(T_{\rm ann})}{4}  \frac{g_\star(T_{\rm ann})}{g_\star(T_*)} \left(\frac{T_{\rm ann}}{T_*}\right)^4 =1 .
 \label{Eq:p(T)}
\end{equation}
This defines $T_*$, and its corresponding lifetime $t_*$. The PBH mass is then given by the mass $M(t_*)$, 
\begin{equation} \label{eq:MPBH}
    M_{\rm PBH} = M(t_*) \simeq \frac{4 \pi}{3}V_{\rm bias} t_*^3 \simeq \frac{2}{[p(T_{\rm ann})]^{3/2}} M(t_{\rm ann}) \simeq \left(\frac{3}{32 \pi}\right)^{1/2} \frac{M_P^3}{V_{\rm bias}^{1/2}} .
\end{equation}
Therefore, inverting Eq.~\eqref{eq:MPBH} and using Eq.~\eqref{eq:ptRadius} one finds,
\begin{equation}
  p(T_{\rm ann})\simeq \frac{t_{\rm ann}^2M_P^4}{M_{\rm PBH}^2}  =
   \frac{90}{32 \pi^3}\frac{1}{g_\star(T_{\rm ann})}  \frac{M_P^6}{T_{\rm ann}^4 M_{\rm PBH}^2} = \frac{0.24}{g_\star (T_{\rm ann})}
    \left(\frac{\rm 10~ eV}{T_{\rm ann}}\right)^4
    \left(\frac{10^{16} M_\odot}{M_{\rm PBH}}\right)^2.
   \label{eq:pann-1}
\end{equation}

Due to effects of deviation from the spherical shape and angular momentum, the probability of forming a PBH at temperature $T$ may be smaller than $p(T)$, say $p(T)^\beta$, with  a real coefficient $\beta \geq 1$. This could account for small deviations from sphericity. A large enough deviation from the spherical shape could prevent the formation of a PBH, since the degree of asymmetry decreases initially but increases in the late stages of the collapse~\cite{Widrow:1989fe}. However, this is unlikely, as shown in the context of the collapse of vacuum bubble produced during inflation~\cite{Deng:2017uwc}. We proceed assuming that for some portion of the walls the asymmetry during collapse is small enough.
So the PBH density at formation is $\rho_{\rm PBH}(T_*) \simeq  p^\beta(T_*) \rho_{\rm wall}(T_*)$,  and the  fraction $f_{\rm PBH}$ of the DM in PBHs is
\begin{equation} \label{eq:PBH fraction}
    f_{\rm PBH}=\frac{\rho_{\rm PBH}(T_*)}{\rho_{\rm CDM}(T_*)}     \simeq p^\beta(T_*)\frac{\rho_{\rm wall}(T_*)}{\rho_{\rm CDM}(T_*)}=
    \frac{\rho_{\rm wall}(T_*)}{\rho_{\rm wall}(T_{\rm ann})}
    \frac{\rho_{\rm wall}(T_{\rm ann})}{\rho_{\rm CDM}(T_*)} ,
\end{equation}
where we used that at the production temperature $p(T_*) =  1 $. The bulk of the energy density of the string-wall network goes into axions at annihilation, thus $\rho_{\rm wall}(T_{\rm ann}) \simeq \rho_a(T_{\rm ann}) = f_{\rm ALP}~ \rho_{\rm CDM}(T_{\rm ann})$, where $f_{\rm ALP}$ is the fraction of the DM in ALPs.
Using Eq.~\eqref{Eq:p(T)} and approximating the evolution of the wall energy density with temperature after annihilation by
\begin{equation}
   \frac{\rho_{\rm wall}(T)}{\rho_{\rm wall}(T_{\rm ann})}=
    \left(\frac{T}{T_{\rm ann}} \right)^\alpha ,
\end{equation}
with the power $\alpha$  extracted from simulations~\cite{Kawasaki:2014sqa}, 
we find that
\begin{equation}
    f_{\rm PBH} \simeq f_{\rm ALP}~ \left[\frac{p(T_{\rm ann})}{4}\right]^{(\alpha-3)/4}~
    \left[ \frac{g_\star(T_{\rm ann})}{g_\star(T_*)}\right]^{(\alpha-3)/4} \frac{g_{s \star}(T_{\rm ann})}{g_{s \star}(T_*)} ,
\end{equation}
since the axion number density redshifts as $T^3$.
Numerical simulations~\cite{Kawasaki:2014sqa} give the two values of times $t(10\%)$ and $t(1\%)$ at which 10\% and 1\% of the string-wall system energy density remains after annihilation has started. Table~VI  and Fig.~4 of Ref.~\cite{Kawasaki:2014sqa} show that the $t(1\%)/t(10\%)$   ratio takes up values close to 2, actually from 1.7 to 1.5, under different assumptions.  This translates into values of the exponent $\alpha$ roughly between 7 and 12.
Hence, $(\alpha-3)/4$ is between 1 and 2. Neglecting the possible change of degrees of freedom between $T_{\rm ann}$ and $T_*$, since these temperatures are close to each other, we find that $f_{\rm PBH}$ is in the range
\begin{equation} 
    f_{\rm PBH} \in \left\{ f_{\rm ALP} \left[\frac{p(T_{\rm ann})}{4}\right], \,  f_{\rm ALP} \left[\frac{p(T_{\rm ann})}{4}\right]^2\right\} .
    \label{eq:fPBH}
\end{equation}

Both $p(T_{\rm ann})$ (as shown in Eq.~\eqref{eq:pann-1})  as well as the relic axion density $\Omega_a h^2$, Eq.~\eqref{Omega-a-walls},  can be written as a function of the annihilation temperature $T_{\rm ann}$ and the PBH mass
$M_{\rm PBH}$,
\begin{equation}
    \Omega_a h^2=\frac{3.5\times 10^7}{g_{s \star}(T_{\rm ann})} \left(\frac{\mathrm{GeV}}{T_{\mathrm{ann}}} \right)^3
    \left(\frac{M_\odot}{M_{\mathrm{PBH}}}\right)^2 .
\end{equation}
Thus we can write $\Omega_a h^2$ as function of $M_{\rm PBH}$ and $p(T_{\rm ann})$,
\begin{equation}
    \Omega_a h^2 = 3.2~[p(T_{\rm ann})]^{3/4}~ \left(\frac{10^{15} M_\odot}{M_{\rm PBH}}\right)^{1/2}~
    \frac{[g_\star(T_{\rm ann})]^{3/4}}{g_{s \star}(T_{\rm ann})} .
\end{equation}
Requiring axions to constitute a fraction $f_{\rm ALP}$ of the DM, $\Omega_a h^2= f_{\rm ALP}\, 0.12$, leads to the following expressions for $p(T_{\rm ann})$ and $M_{\rm PBH}$,
\begin{equation} \label{pann-MPBH}
    p(T_{\rm ann}) = 0.27~ f_{\rm ALP}^{4/3} \left(\frac{M_{\rm PBH}}{10^{17} M_\odot}\right)^{2/3}
    \frac{[g_{s \star}(T_{\rm ann})]^{4/3}}{g_\star(T_{\rm ann})} .
\end{equation}
\begin{equation} \label{MPBH-Tann}
    \frac{M_{\rm PBH}}{M_\odot}= \frac{1.7\times 10^4}{[g_{s \star}(T_{\rm ann})]^{1/2}}\left(\frac{\rm GeV}{T_{\rm ann}}\right)^{3/2}f_{\rm ALP}^{-1/2}.
\end{equation}
Therefore, an upper limit on $f_{\rm ALP}$ implies an upper limit on the fraction $f_{\rm PBH}$ of the DM in PBHs, as per  Eq.~\eqref{eq:fPBH} and a lower limit on the PBH mass. 

We show the density of PBHs as $f_{\rm PBH}=\Omega_{\rm PBH}/\Omega_{\rm DM}$ in Fig.~\ref{fig:PBH} for respectively $f_{\rm ALP}=1$ (upper panel) and $f_{\rm ALP}=0.1$ (lower panel). The bounds (regions contoured by solid lines), taken from Ref.~\cite{Carr:2020pbh} and references therein, include millilensing of compact radio
sources (RS), dynamical limits from disruption of wide binaries (WB) and globular clusters (GC), heating of stars in the
Galactic disk (DH), dynamical
friction constraints (DF), disruption of galaxies (G), and the CMB dipole bound (CMB). The incredulity limit
(IL) is found assuming the presence of at least one PBH per Hubble volume. Projected realistic (D1) and optimistic (D2) discovery limits through $\mu$-distortion of the CMB spectrum correspond to the  regions contoured by dashed lines and are taken from Ref.~\cite{Deng:2021edw}.
For a fixed ALP density, our predicted PBH population could fall within the indicated black funnel corresponding to the range in Eq.~\eqref{eq:fPBH} where $p_{\rm ann}$ as function of $M_{\rm PBH}$ is given by  Eq.~\eqref{pann-MPBH}. The green vertical lines within each funnel show the PBH mass corresponding to each indicated value of the annihilation temperature given by Eq.~\eqref{MPBH-Tann}.

Having $f_{\rm ALP}=1$ (upper panel) is probably only possible with high enough annihilation temperatures, e.g. $T_{\rm ann}\gtrsim \mathcal{O}(\rm keV)$, due to structure formation contraints. Thus we cut the allowed funnel region to the right of the $T_{\rm ann} =$ 5 keV vertical line. 
In this case, the collapse of the closed spherical walls could produce ``stupendously'' large black holes (as named in Ref.~\cite{Carr:2020erq}) with mass up to $10^{12} M_\odot$. If the annihilation happens at temperatures above 500 keV, it could produce supermassive black holes (SMBHs) with mass up to $10^9 M_\odot$ as those  found at the center of large galaxies. These black holes are observed at large redshift~\cite{Haiman:2000ky,2021ApJ...907L...1W}, and their production through accretion and mergers of smaller black holes from Pop III stars  is generally complicated, and requires either a large initial seed or an increased growth rate (for a review see Ref.~\cite{Inayoshi:2019fun}), though turbulent cold flows might help~\cite{Latif}. Therefore, the existence of SMBHs could be a tantalizing hint of physics beyond the SM at work. Other mechanisms have been recently proposed to produce SMBHs, and for completeness we briefly comment on their similarities and differences with ours.

A SMBH production mechanism relying on a first order phase transition has been recently advanced~\cite{Davoudiasl:2021ijv}, which is in tension with heavy quasar superradiance bounds~\cite{Unal:2020jiy}. Our model instead avoids these bounds for large enough $m_a$. 
An alternative scenario invoking physics beyond the SM is that of the gravothermal collapse of self-interacting  DM halos   (see e.g.~\cite{Balberg:2001qg,Balberg:2002ue,Pollack:2014rja}). However, since 
the cross sections needed for the gravothermal collapse are ruled out by observations
of galaxy cluster collisions, more complicated models need to be built, including totally dissipative DM (i.e., with ``hit-and-stick" collisions~\cite{Xiao:2021ftk}) or mixed DM (in which the DM has two components, and one component has a large self-interaction cross section, e.g.~\cite{Choquette:2018lvq}). Finally, SMBHs might be produced through PBH mergers~\cite{Duechting:2004dk}, and collapse of domain walls~\cite{Khlopov:2004sc} or vacuum bubbles~\cite{Deng:2017uwc,Kusenko:2020pcg} nucleated during inflation. We stress that our scenario is particularly economical: both the ALP DM density and the PBH population depend only on two macroscopic parameters, $\sigma$ and $V_{\rm bias}$. The expected mass of the SMBHs can reach up to $10^9\, M_\odot$, which  corresponds through Eq.~\eqref{MPBH-Tann} to annihilation temperatures larger than $0.66 f_{\rm ALP}^{-1/3}$ MeV, and through Eq.~\eqref{eq:MPBH} to a lower limit on $V_{\rm bias}$. Equation~\eqref{Omega-a-walls} shows that these limits translate to limits on $\sigma$ for each value of  $f_{ALP}$. However, these limits cannot be directly translated into an ALP mass bound since the dynamics of the string-wall network depends on 2 macroscopic parameters, $\sigma$ and $V_{\rm bias}$, but there are 3 parameters in the Lagrangian of Eq.~\eqref{eq:potential}, $\epsilon_b$,  $v$, and $V$. 

The lower panel of Fig.~\ref{fig:PBH} presents our results for $f_{\rm ALP}=0.1$, which is compatible with  $T_{\rm ann} \gtrsim 5\,\rm eV$, since structure formation bounds do not apply to a subdominant component of the DM. So we cut the right portion of the funnels accordingly.  In this case
there could be production of ``stupendously'' large black holes~\cite{Carr:2020erq,Deng:2021edw} with mass up to
$10^{17}\, M_\odot$ and density as large as $10^{-4}$ of the DM density. These black holes necessarily cannot reside in galaxies~\cite{Carr:2020erq}. It is not clear if these PBHs can exist, and we remain agnostic about the mechanism that could segregate them outside of galaxies. They will be potentially probed by measurements of the CMB spectral $\mu$-distortions in future  experiments~\cite{Deng:2021edw}. If the annihilation temperature is above 500 keV, SMBHs would again be produced, although with a lower density than for larger $f_{\rm ALP}$ values.

\section{Bias from Planck scale physics}

Until now, we have been agnostic about the physics introducing the small bias in the potential of Eq.~\eqref{eq:potential}. Quantum gravity effects should lead to the violation of global symmetries, since classical black holes have no global charge~\cite{Harlow:2018tng}. Therefore, when there is a global $U(1)$ symmetry, one can expect a series of terms with dimension $n+4$ suppressed by powers of the Planck mass,
\begin{equation}\label{eq:planckoperators}
    \mathcal{L}_I^{\rm QG}\sim \sum_{n=1}^\infty g_P^{(n)}\phi^3 (\phi e^{-i\delta}+\phi^*e^{i\delta}) \frac{\phi^n}{m_P^n}.
\end{equation}
Here we would like to associate our bias term, given by the third term of Eq.~\eqref{eq:potential}, with the Planck scale-suppressed operators of Eq.~\eqref{eq:planckoperators}.

 In the case of the QCD axion, similar to the more general ALP setting we are analyzing, the bias is necessary to get rid of the string-wall system for $N>1$ in the post-inflationary scenario~\cite{Sikivie:1982qv}.  On the other hand, similar operators can be dangerous and spoil the axion solution to the strong CP problem~\cite{Georgi:1981pu}, introducing the so-called ``axion quality problem"~\cite{Irastorza:2018dyq,DiLuzio:2020wdo}. For example, Planck-suppressed operators with dimension $n+4<10$ can spoil the axion solution of the strong CP problem. Therefore, one needs $n+4\geq 10$. Moreover, operators with dimension 10, 11, and 12 annihilate the string-wall network efficiently~\cite{Barr:1992qq} (see also~\cite{Kamionkowski:1992mf}).

\begin{figure}  
  \centering
  \includegraphics[width=0.7\linewidth]{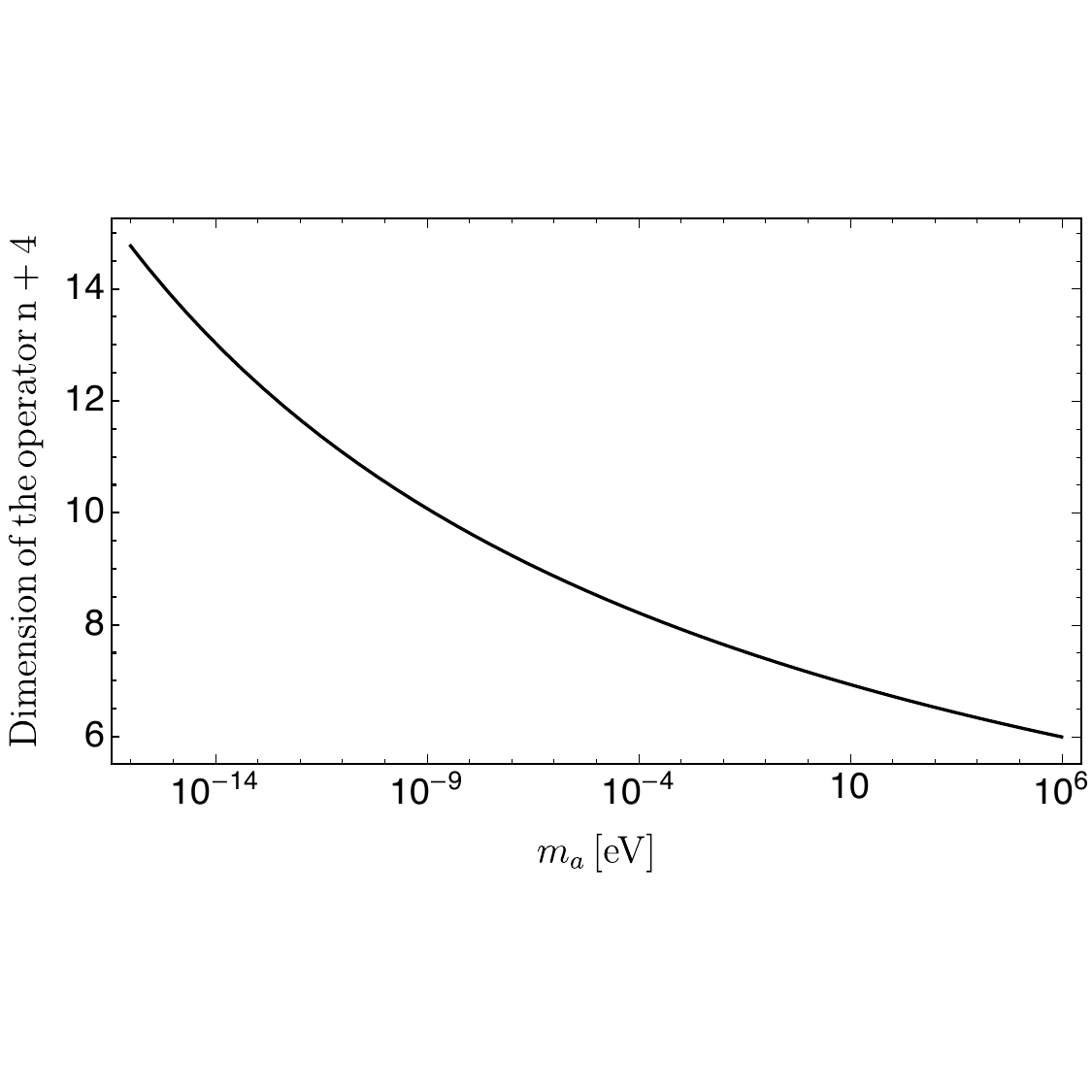}
  \vspace{-4em}
\caption{Planck-scale operator dimension needed to have observable GWs, according to Eq.~\eqref{eq:planck}.}
\label{fig:planckscale}
\end{figure}

To see how this plays out in the more general ALP scenario, let us write the bias as produced by a Planck-suppressed operator, which is given by
\begin{equation}
V_{\rm bias}=\epsilon_b v^4=g_P V^4\frac{V^n}{m_{P}^n},
\end{equation}
where $|\phi|\sim V$ after the spontaneous symmetry breaking.
We can also assume $g_P=\mathcal{O}(1)$ (see however Ref.~\cite{Kallosh:1995hi}), so that we find
\begin{equation}
   \log \epsilon_b \frac{v^4}{V^4}=n\log \frac{V}{m_{P}} 
   \label{eq:planck}.
\end{equation}
Counterintuitively, although the production of heavier ALPs requires small biases to produce observable GWs, we find that the dimension $n$ required decreases with $m_a$.  Using Eqs.~\eqref{V-scaling-window} and~\eqref{eps-scaling-window} to write Eq.~\eqref{eq:planck} as a function of $m_a$ only, we find the result shown in Fig.~\ref{fig:planckscale}. For example, the dimension of the operator needs to be $n+4\simeq 6$ for the heaviest ALP mass we consider (about MeV) and $n+4\simeq 15$ for $m_a\simeq 10^{-16}\,\rm eV$. Fine tuning of parameters is necessary to suppress terms with smaller $n$ when a certain value of $n$ is required.  Therefore, as shown in Fig.~\ref{fig:planckscale}, we see that the production of heavier ALPs needs somewhat less fine-tuning to be observable through GWs. 

\section{Concluding remarks}

We have presented new windows on ALP models in case of catastrogenesis which takes advantage of the fast progress expected in GW detection and in the understanding of the largest black holes in the Universe. 

Axion-like particles are a bosonic DM candidate associated with a global $U(1)$ symmetry which is spontaneously and also explicitly broken. When the spontaneous breaking happens after inflation, as we have assumed, and the explicit breaking yields a number $N>1$ of almost degenerate vacua,  the formation of a stable cosmic string-wall network ensues.
In this case a small energy difference, a bias, between the vacua is needed to avoid the unacceptable cosmological consequences of walls dominating the energy density of the Universe. The volume energy density due to the bias eventually becomes equal to and then dominant over the surface energy density of the walls, at which point it drives
the string-wall system to annihilation.

We have shown that if the bias is such that the annihilation happens shortly before recombination at 5~eV~$\lesssim T_{\mathrm{ann}}\lesssim 10^{2}$~eV, GWs can potentially be detected by future CMB probes and astrometry measurements for ALPs with mass from $10^{-16}$ to $10^{6}$~eV. Structure formation limits impose that these ALPs could constitute only a fraction, possibly up to 0.3,  of the DM. Additionally, after the process of annihilation starts, some closed domain walls could collapse into  PBHs in the supermassive to ``stupendously massive" range.  If ALPs constitute 0.1 of the DM, PBHs of mass up to $10^{17} M_\odot$ and with abundance as large as $10^{-4}$ of the DM density could be produced. These could be probed by future astronomical measurements (see the lower panel of Fig.~\ref{fig:PBH}). For annihilation temperatures larger than a few keV ALPs could constitute the whole of the DM, and for temperatures above 500 keV, supermassive black holes  as those found in the
center of large galaxies could be produced.

Our model therefore potentially links two long standing problems, the nature of the DM and the origin of the largest black holes in the Universe, in an economical manner.

\smallskip
\acknowledgments
{We  thank Josh Ruderman and Tracy Slatyer for useful suggestions. The work of GG and EV was supported in part by the U.S. Department of Energy (DOE) Grant No. DE-SC0009937.}
  \appendix

\section{Additional limits}
\label{tftapp}

The following limits do not affect the regions of interest we presented, but we include them for completeness. The string-wall system would be present during Big Bang Nucleosynthesis, thus it contributes to the effective number of neutrino species~\cite{Hiramatsu:2011eswn}, 
\begin{equation}
    \Delta N_{\rm{eff}}^{\rm walls} \simeq  2.1\times 10^{-21} \left(\frac{f_\sigma m_a}{N^2 {\rm eV}}\right)
 \left(\frac{V}{\rm GeV}\right)^2.
\end{equation}
In the range of  interest of our parameters is always much smaller than present upper limits (close to  0.5~\cite{Aghanim:2018eyx}).

For self-consistency  walls must form before annihilating, i.e. $T_{\rm ann}<T_w$,  which implies a weak upper limit on the bias,
\begin{equation}
    \epsilon_b < 0.5 f_\sigma\left[\frac{g_\star(T_{\rm ann})}{g_\star(T_w)}\right]^{1/2}.
\end{equation}

The string-wall system would eventually dominate the energy density of the Universe at a temperature $T_{wd}$ (assuming radiation domination),
\begin{equation}
    T_{wd}=\left[\frac{40}{\pi^2 g_*(T_{wd})}\right]^{1/4} \left(\frac{\sigma}{m_P}\right)^{1/2}= \frac{3.4 \times 10^{-14}\, {\rm GeV}}{[g_\star(T_{wd})]^{1/4}}
    \left(\frac{V}{N~ {\rm GeV}}\right)
  \left(\frac{f_\sigma ~m_a}{\rm eV}\right)^{1/2}.
\end{equation}
We require that the string-wall system annihilated before, i.e. $T_{\rm ann}>T_{wd}$, which  implies a lower limit on the bias,
\begin{equation}
    \frac{V_{\rm bias}}{{\rm GeV}^4} > 2.4 \times 10^{-37} \left(\frac{\sigma}{{\rm GeV}^3}\right)^2  \left[\frac{g_\ast(T_{\rm ann})}{g_\star(T_{wd})}\right]^{1/2},
\end{equation}
or 
\begin{equation}
    \epsilon_b > 2.39 \times 10^{-37}  \left[\frac{g_\ast(T_{\rm ann})}{g_\star(T_{wd})}\right]^{1/2}
    \left(\frac{f_\sigma~V}{N~{\rm GeV}}\right)^2.
\end{equation}

\bibliographystyle{bibi}
\bibliography{bibliography}

\providecommand{\href}[2]{#2}\begingroup\raggedright\begin{thebibliography}{100}

\bibitem{Vilenkin:2000jqa}
A.~Vilenkin and E.~P.~S. Shellard, \emph{{Cosmic Strings and Other Topological
  Defects}}. Cambridge University Press, 7, 2000.

\bibitem{Maggiore:1900zz}
M.~Maggiore, \emph{{Gravitational Waves. Vol. 1: Theory and Experiments}},
  Oxford Master Series in Physics. Oxford University Press, 2007.

\bibitem{Maggiore:2018sht}
M.~Maggiore, \emph{{Gravitational Waves. Vol. 2: Astrophysics and Cosmology}}.
  Oxford University Press, 3, 2018.

\bibitem{Sathyaprakash:2009xs}
B.~S. Sathyaprakash and B.~F. Schutz, \emph{{Physics, Astrophysics and
  Cosmology with Gravitational Waves}},
  \href{https://doi.org/10.12942/lrr-2009-2}{\emph{Living Rev. Rel.} {\bfseries
  12} (2009) 2} [\href{https://arxiv.org/abs/0903.0338}{{\ttfamily
  0903.0338}}].

\bibitem{Barack:2018yly}
L.~Barack et~al., \emph{{Black holes, gravitational waves and fundamental
  physics: a roadmap}},
  \href{https://doi.org/10.1088/1361-6382/ab0587}{\emph{Class. Quant. Grav.}
  {\bfseries 36} (2019) 143001}
  [\href{https://arxiv.org/abs/1806.05195}{{\ttfamily 1806.05195}}].

\bibitem{Gelmini:2021yzu}
G.~B. Gelmini, A.~Simpson and E.~Vitagliano, \emph{{Gravitational waves from
  axionlike particle cosmic string-wall networks}},
  \href{https://doi.org/10.1103/PhysRevD.104.L061301}{\emph{Phys. Rev. D}
  {\bfseries 104} (2021) 061301}
  [\href{https://arxiv.org/abs/2103.07625}{{\ttfamily 2103.07625}}].

\bibitem{Namikawa:2019tax}
T.~Namikawa, S.~Saga, D.~Yamauchi and A.~Taruya, \emph{{CMB Constraints on the
  Stochastic Gravitational-Wave Background at Mpc scales}},
  \href{https://doi.org/10.1103/PhysRevD.100.021303}{\emph{Phys. Rev. D}
  {\bfseries 100} (2019) 021303}
  [\href{https://arxiv.org/abs/1904.02115}{{\ttfamily 1904.02115}}].

\bibitem{Peccei:1977hh}
R.~D. Peccei and H.~R. Quinn, \emph{{CP Conservation in the Presence of
  Instantons}}, \href{https://doi.org/10.1103/PhysRevLett.38.1440}{\emph{Phys.
  Rev. Lett.} {\bfseries 38} (1977) 1440}.

\bibitem{Weinberg:1977ma}
S.~Weinberg, \emph{{A New Light Boson?}},
  \href{https://doi.org/10.1103/PhysRevLett.40.223}{\emph{Phys. Rev. Lett.}
  {\bfseries 40} (1978) 223}.

\bibitem{Wilczek:1977pj}
F.~Wilczek, \emph{{Problem of Strong $P$ and $T$ Invariance in the Presence of
  Instantons}}, \href{https://doi.org/10.1103/PhysRevLett.40.279}{\emph{Phys.
  Rev. Lett.} {\bfseries 40} (1978) 279}.

\bibitem{Kim:1979if}
J.~E. Kim, \emph{{Weak Interaction Singlet and Strong CP Invariance}},
  \href{https://doi.org/10.1103/PhysRevLett.43.103}{\emph{Phys. Rev. Lett.}
  {\bfseries 43} (1979) 103}.

\bibitem{Shifman:1979if}
M.~A. Shifman, A.~I. Vainshtein and V.~I. Zakharov, \emph{{Can Confinement
  Ensure Natural CP Invariance of Strong Interactions?}},
  \href{https://doi.org/10.1016/0550-3213(80)90209-6}{\emph{Nucl. Phys. B}
  {\bfseries 166} (1980) 493}.

\bibitem{Dine:1981rt}
M.~Dine, W.~Fischler and M.~Srednicki, \emph{{A Simple Solution to the Strong
  CP Problem with a Harmless Axion}},
  \href{https://doi.org/10.1016/0370-2693(81)90590-6}{\emph{Phys. Lett. B}
  {\bfseries 104} (1981) 199}.

\bibitem{Zhitnitsky:1980tq}
A.~R. Zhitnitsky, \emph{{On Possible Suppression of the Axion Hadron
  Interactions. (In Russian)}}, {\emph{Sov. J. Nucl. Phys.} {\bfseries 31}
  (1980) 260}.

\bibitem{Chikashige:1980ui}
Y.~Chikashige, R.~N. Mohapatra and R.~D. Peccei, \emph{{Are There Real
  Goldstone Bosons Associated with Broken Lepton Number?}},
  \href{https://doi.org/10.1016/0370-2693(81)90011-3}{\emph{Phys. Lett. B}
  {\bfseries 98} (1981) 265}.

\bibitem{Gelmini:1980re}
G.~B. Gelmini and M.~Roncadelli, \emph{{Left-Handed Neutrino Mass Scale and
  Spontaneously Broken Lepton Number}},
  \href{https://doi.org/10.1016/0370-2693(81)90559-1}{\emph{Phys. Lett. B}
  {\bfseries 99} (1981) 411}.

\bibitem{Wilczek:1982rv}
F.~Wilczek, \emph{{Axions and Family Symmetry Breaking}},
  \href{https://doi.org/10.1103/PhysRevLett.49.1549}{\emph{Phys. Rev. Lett.}
  {\bfseries 49} (1982) 1549}.

\bibitem{Reiss:1982sq}
D.~B. Reiss, \emph{{Can the Family Group Be a Global Symmetry?}},
  \href{https://doi.org/10.1016/0370-2693(82)90647-5}{\emph{Phys. Lett. B}
  {\bfseries 115} (1982) 217}.

\bibitem{Gelmini:1982zz}
G.~B. Gelmini, S.~Nussinov and T.~Yanagida, \emph{{Does Nature Like
  Nambu-Goldstone Bosons?}},
  \href{https://doi.org/10.1016/0550-3213(83)90426-1}{\emph{Nucl. Phys. B}
  {\bfseries 219} (1983) 31}.

\bibitem{Svrcek:2006yi}
P.~Svrcek and E.~Witten, \emph{{Axions In String Theory}},
  \href{https://doi.org/10.1088/1126-6708/2006/06/051}{\emph{JHEP} {\bfseries
  06} (2006) 051} [\href{https://arxiv.org/abs/hep-th/0605206}{{\ttfamily
  hep-th/0605206}}].

\bibitem{Arvanitaki:2009fg}
A.~Arvanitaki, S.~Dimopoulos, S.~Dubovsky, N.~Kaloper and J.~March-Russell,
  \emph{{String Axiverse}},
  \href{https://doi.org/10.1103/PhysRevD.81.123530}{\emph{Phys. Rev. D}
  {\bfseries 81} (2010) 123530}
  [\href{https://arxiv.org/abs/0905.4720}{{\ttfamily 0905.4720}}].

\bibitem{Acharya:2010zx}
B.~S. Acharya, K.~Bobkov and P.~Kumar, \emph{{An M Theory Solution to the
  Strong CP Problem and Constraints on the Axiverse}},
  \href{https://doi.org/10.1007/JHEP11(2010)105}{\emph{JHEP} {\bfseries 11}
  (2010) 105} [\href{https://arxiv.org/abs/1004.5138}{{\ttfamily 1004.5138}}].

\bibitem{Dine:2010cr}
M.~Dine, G.~Festuccia, J.~Kehayias and W.~Wu, \emph{{Axions in the Landscape
  and String Theory}},
  \href{https://doi.org/10.1007/JHEP01(2011)012}{\emph{JHEP} {\bfseries 01}
  (2011) 012} [\href{https://arxiv.org/abs/1010.4803}{{\ttfamily 1010.4803}}].

\bibitem{Jaeckel:2010ni}
J.~Jaeckel and A.~Ringwald, \emph{{The Low-Energy Frontier of Particle
  Physics}},
  \href{https://doi.org/10.1146/annurev.nucl.012809.104433}{\emph{Ann. Rev.
  Nucl. Part. Sci.} {\bfseries 60} (2010) 405}
  [\href{https://arxiv.org/abs/1002.0329}{{\ttfamily 1002.0329}}].

\bibitem{Vilenkin:1984ib}
A.~Vilenkin, \emph{{Cosmic Strings and Domain Walls}},
  \href{https://doi.org/10.1016/0370-1573(85)90033-X}{\emph{Phys. Rept.}
  {\bfseries 121} (1985) 263}.

\bibitem{Zeldovich:1974uw}
Y.~Zeldovich, I.~Kobzarev and L.~Okun, \emph{{Cosmological Consequences of the
  Spontaneous Breakdown of Discrete Symmetry}}, {\emph{Zh. Eksp. Teor. Fiz.}
  {\bfseries 67} (1974) 3}.

\bibitem{Sikivie:1982qv}
P.~Sikivie, \emph{{Of Axions, Domain Walls and the Early Universe}},
  \href{https://doi.org/10.1103/PhysRevLett.48.1156}{\emph{Phys. Rev. Lett.}
  {\bfseries 48} (1982) 1156}.

\bibitem{Chang:1998bq}
S.~Chang, C.~Hagmann and P.~Sikivie, \emph{{Axions from wall decay}},
  \href{https://doi.org/10.1016/S0920-5632(98)00510-6}{\emph{Nucl. Phys. B
  Proc. Suppl.} {\bfseries 72} (1999) 99}
  [\href{https://arxiv.org/abs/hep-ph/9808302}{{\ttfamily hep-ph/9808302}}].

\bibitem{Gelmini:1988sf}
G.~B. Gelmini, M.~Gleiser and E.~W. Kolb, \emph{{Cosmology of Biased Discrete
  Symmetry Breaking}},
  \href{https://doi.org/10.1103/PhysRevD.39.1558}{\emph{Phys. Rev. D}
  {\bfseries 39} (1989) 1558}.

\bibitem{Chang:2019mza}
C.-F. Chang and Y.~Cui, \emph{{Stochastic Gravitational Wave Background from
  Global Cosmic Strings}},
  \href{https://doi.org/10.1016/j.dark.2020.100604}{\emph{Phys. Dark Univ.}
  {\bfseries 29} (2020) 100604}
  [\href{https://arxiv.org/abs/1910.04781}{{\ttfamily 1910.04781}}].

\bibitem{Gorghetto:2021fsn}
M.~Gorghetto, E.~Hardy and H.~Nicolaescu, \emph{{Observing invisible axions
  with gravitational waves}},
  \href{https://doi.org/10.1088/1475-7516/2021/06/034}{\emph{JCAP} {\bfseries
  06} (2021) 034} [\href{https://arxiv.org/abs/2101.11007}{{\ttfamily
  2101.11007}}].

\bibitem{Higaki:2016jjh}
T.~Higaki, K.~S. Jeong, N.~Kitajima, T.~Sekiguchi and F.~Takahashi,
  \emph{{Topological Defects and nano-Hz Gravitational Waves in Aligned Axion
  Models}}, \href{https://doi.org/10.1007/JHEP08(2016)044}{\emph{JHEP}
  {\bfseries 08} (2016) 044}
  [\href{https://arxiv.org/abs/1606.05552}{{\ttfamily 1606.05552}}].

\bibitem{ZambujalFerreira:2021cte}
R.~Zambujal~Ferreira, A.~Notari, O.~Pujol\`as and F.~Rompineve, \emph{{High
  Quality QCD Axion at Gravitational Wave Observatories}},
  \href{https://doi.org/10.1103/PhysRevLett.128.141101}{\emph{Phys. Rev. Lett.}
  {\bfseries 128} (2022) 141101}
  [\href{https://arxiv.org/abs/2107.07542}{{\ttfamily 2107.07542}}].

\bibitem{Takahashi:2020tqv}
F.~Takahashi and W.~Yin, \emph{{Kilobyte Cosmic Birefringence from ALP Domain
  Walls}}, \href{https://doi.org/10.1088/1475-7516/2021/04/007}{\emph{JCAP}
  {\bfseries 04} (2021) 007}
  [\href{https://arxiv.org/abs/2012.11576}{{\ttfamily 2012.11576}}].

\bibitem{Kitajima:2022jzz}
N.~Kitajima, F.~Kozai, F.~Takahashi and W.~Yin, \emph{{Power spectrum of
  domain-wall network and its implications for isotropic and anisotropic cosmic
  birefringence}},  \href{https://arxiv.org/abs/2205.05083}{{\ttfamily
  2205.05083}}.

\bibitem{Ferrer:2019pbh}
F.~Ferrer, E.~Masso, G.~Panico, O.~Pujolas and F.~Rompineve, \emph{{Primordial
  Black Holes from the QCD Axion}},
  \href{https://doi.org/10.1103/PhysRevLett.122.10301}{\emph{Phys. Rev. Lett.}
  {\bfseries 122} (2019) 10301}
  [\href{https://arxiv.org/abs/1807.01707}{{\ttfamily 1807.01707}}].

\bibitem{Pagano:2015hma}
L.~Pagano, L.~Salvati and A.~Melchiorri, \emph{{New constraints on primordial
  gravitational waves from Planck 2015}},
  \href{https://doi.org/10.1016/j.physletb.2016.07.078}{\emph{Phys. Lett. B}
  {\bfseries 760} (2016) 823}
  [\href{https://arxiv.org/abs/1508.02393}{{\ttfamily 1508.02393}}].

\bibitem{Laureijs:2011gra}
{\scshape EUCLID} Collaboration, R.~Laureijs et~al., \emph{{Euclid Definition
  Study Report}},  \href{https://arxiv.org/abs/1110.3193}{{\ttfamily
  1110.3193}}.

\bibitem{Darling:2018hmc}
J.~Darling, A.~E. Truebenbach and J.~Paine, \emph{{Astrometric Limits on the
  Stochastic Gravitational Wave Background}},
  \href{https://doi.org/10.3847/1538-4357/aac772}{\emph{Astrophys. J.}
  {\bfseries 861} (2018) 113}
  [\href{https://arxiv.org/abs/1804.06986}{{\ttfamily 1804.06986}}].

\bibitem{Arvanitaki:2019rax}
A.~Arvanitaki, S.~Dimopoulos, M.~Galanis, L.~Lehner, J.~O. Thompson and
  K.~Van~Tilburg, \emph{{Large-misalignment mechanism for the formation of
  compact axion structures: Signatures from the QCD axion to fuzzy dark
  matter}}, \href{https://doi.org/10.1103/PhysRevD.101.083014}{\emph{Phys. Rev.
  D} {\bfseries 101} (2020) 083014}
  [\href{https://arxiv.org/abs/1909.11665}{{\ttfamily 1909.11665}}].

\bibitem{Carr:2020erq}
B.~Carr, F.~Kuhnel and L.~Visinelli, \emph{{Constraints on Stupendously Large
  Black Holes}}, \href{https://doi.org/10.1093/mnras/staa3651}{\emph{Mon. Not.
  Roy. Astron. Soc.} {\bfseries 501} (2021) 2029}
  [\href{https://arxiv.org/abs/2008.08077}{{\ttfamily 2008.08077}}].

\bibitem{Deng:2021edw}
H.~Deng, \emph{{\ensuremath{\mu}-distortion around stupendously large
  primordial black holes}},
  \href{https://doi.org/10.1088/1475-7516/2021/11/054}{\emph{JCAP} {\bfseries
  11} (2021) 054} [\href{https://arxiv.org/abs/2106.09817}{{\ttfamily
  2106.09817}}].

\bibitem{Hertzberg:2008wr}
M.~P. Hertzberg, M.~Tegmark and F.~Wilczek, \emph{{Axion Cosmology and the
  Energy Scale of Inflation}},
  \href{https://doi.org/10.1103/PhysRevD.78.083507}{\emph{Phys. Rev. D}
  {\bfseries 78} (2008) 083507}
  [\href{https://arxiv.org/abs/0807.1726}{{\ttfamily 0807.1726}}].

\bibitem{Aghanim:2018eyx}
{\scshape Planck} Collaboration, N.~Aghanim et~al., \emph{{Planck 2018 results.
  VI. Cosmological parameters}},
  \href{https://doi.org/10.1051/0004-6361/201833910}{\emph{Astron. Astrophys.}
  {\bfseries 641} (2020) A6}
  [\href{https://arxiv.org/abs/1807.06209}{{\ttfamily 1807.06209}}].

\bibitem{Saikawa:2018rcs}
K.~Saikawa and S.~Shirai, \emph{{Primordial gravitational waves, precisely: The
  role of thermodynamics in the Standard Model}},
  \href{https://doi.org/10.1088/1475-7516/2018/05/035}{\emph{JCAP} {\bfseries
  05} (2018) 035} [\href{https://arxiv.org/abs/1803.01038}{{\ttfamily
  1803.01038}}].

\bibitem{Chang:1998tb}
S.~Chang, C.~Hagmann and P.~Sikivie, \emph{{Studies of the motion and decay of
  axion walls bounded by strings}},
  \href{https://doi.org/10.1103/PhysRevD.59.023505}{\emph{Phys. Rev. D}
  {\bfseries 59} (1999) 023505}
  [\href{https://arxiv.org/abs/hep-ph/9807374}{{\ttfamily hep-ph/9807374}}].

\bibitem{Hiramatsu:2010yz}
T.~Hiramatsu, M.~Kawasaki and K.~Saikawa, \emph{{Gravitational Waves from
  Collapsing Domain Walls}},
  \href{https://doi.org/10.1088/1475-7516/2010/05/032}{\emph{JCAP} {\bfseries
  05} (2010) 032} [\href{https://arxiv.org/abs/1002.1555}{{\ttfamily
  1002.1555}}].

\bibitem{Hiramatsu:2013qaa}
T.~Hiramatsu, M.~Kawasaki and K.~Saikawa, \emph{{On the estimation of
  gravitational wave spectrum from cosmic domain walls}},
  \href{https://doi.org/10.1088/1475-7516/2014/02/031}{\emph{JCAP} {\bfseries
  02} (2014) 031} [\href{https://arxiv.org/abs/1309.5001}{{\ttfamily
  1309.5001}}].

\bibitem{Kawasaki:2011vv}
M.~Kawasaki and K.~Saikawa, \emph{{Study of gravitational radiation from cosmic
  domain walls}},
  \href{https://doi.org/10.1088/1475-7516/2011/09/008}{\emph{JCAP} {\bfseries
  09} (2011) 008} [\href{https://arxiv.org/abs/1102.5628}{{\ttfamily
  1102.5628}}].

\bibitem{Hiramatsu:2012sc}
T.~Hiramatsu, M.~Kawasaki, K.~Saikawa and T.~Sekiguchi, \emph{{Axion cosmology
  with long-lived domain walls}},
  \href{https://doi.org/10.1088/1475-7516/2013/01/001}{\emph{JCAP} {\bfseries
  01} (2013) 001} [\href{https://arxiv.org/abs/1207.3166}{{\ttfamily
  1207.3166}}].

\bibitem{Gelmini:2020bqg}
G.~B. Gelmini, S.~Pascoli, E.~Vitagliano and Y.-L. Zhou, \emph{{Gravitational
  wave signatures from discrete flavor symmetries}},
  \href{https://doi.org/10.1088/1475-7516/2021/02/032}{\emph{JCAP} {\bfseries
  02} (2021) 032} [\href{https://arxiv.org/abs/2009.01903}{{\ttfamily
  2009.01903}}].

\bibitem{Caprini:2009fx}
C.~Caprini, R.~Durrer, T.~Konstandin and G.~Servant, \emph{{General Properties
  of the Gravitational Wave Spectrum from Phase Transitions}},
  \href{https://doi.org/10.1103/PhysRevD.79.083519}{\emph{Phys. Rev. D}
  {\bfseries 79} (2009) 083519}
  [\href{https://arxiv.org/abs/0901.1661}{{\ttfamily 0901.1661}}].

\bibitem{Kamionkowski:1999qc}
M.~Kamionkowski and A.~Kosowsky, \emph{{The Cosmic microwave background and
  particle physics}},
  \href{https://doi.org/10.1146/annurev.nucl.49.1.77}{\emph{Ann. Rev. Nucl.
  Part. Sci.} {\bfseries 49} (1999) 77}
  [\href{https://arxiv.org/abs/astro-ph/9904108}{{\ttfamily
  astro-ph/9904108}}].

\bibitem{Smith:2005mm}
T.~L. Smith, M.~Kamionkowski and A.~Cooray, \emph{{Direct detection of the
  inflationary gravitational wave background}},
  \href{https://doi.org/10.1103/PhysRevD.73.023504}{\emph{Phys. Rev. D}
  {\bfseries 73} (2006) 023504}
  [\href{https://arxiv.org/abs/astro-ph/0506422}{{\ttfamily
  astro-ph/0506422}}].

\bibitem{Clarke:2020bil}
T.~J. Clarke, E.~J. Copeland and A.~Moss, \emph{{Constraints on primordial
  gravitational waves from the Cosmic Microwave Background}},
  \href{https://doi.org/10.1088/1475-7516/2020/10/002}{\emph{JCAP} {\bfseries
  10} (2020) 002} [\href{https://arxiv.org/abs/2004.11396}{{\ttfamily
  2004.11396}}].

\bibitem{Lasky:2015lej}
P.~D. Lasky et~al., \emph{{Gravitational-wave cosmology across 29 decades in
  frequency}}, \href{https://doi.org/10.1103/PhysRevX.6.011035}{\emph{Phys.
  Rev. X} {\bfseries 6} (2016) 011035}
  [\href{https://arxiv.org/abs/1511.05994}{{\ttfamily 1511.05994}}].

\bibitem{Campeti:2020xwn}
P.~Campeti, E.~Komatsu, D.~Poletti and C.~Baccigalupi, \emph{{Measuring the
  spectrum of primordial gravitational waves with CMB, PTA and Laser
  Interferometers}},
  \href{https://doi.org/10.1088/1475-7516/2021/01/012}{\emph{JCAP} {\bfseries
  01} (2021) 012} [\href{https://arxiv.org/abs/2007.04241}{{\ttfamily
  2007.04241}}].

\bibitem{Ade:2018gkx}
{\scshape BICEP2, Keck Array} Collaboration, P.~A.~R. Ade et~al., \emph{{BICEP2
  / Keck Array x: Constraints on Primordial Gravitational Waves using Planck,
  WMAP, and New BICEP2/Keck Observations through the 2015 Season}},
  \href{https://doi.org/10.1103/PhysRevLett.121.221301}{\emph{Phys. Rev. Lett.}
  {\bfseries 121} (2018) 221301}
  [\href{https://arxiv.org/abs/1810.05216}{{\ttfamily 1810.05216}}].

\bibitem{Matsumura:2013aja}
T.~Matsumura et~al., \emph{{Mission design of LiteBIRD}},
  \href{https://doi.org/10.1007/s10909-013-0996-1}{\emph{J. Low Temp. Phys.}
  {\bfseries 176} (2014) 733}
  [\href{https://arxiv.org/abs/1311.2847}{{\ttfamily 1311.2847}}].

\bibitem{Hanany:2019lle}
{\scshape NASA PICO} Collaboration, S.~Hanany et~al., \emph{{PICO: Probe of
  Inflation and Cosmic Origins}},
  \href{https://arxiv.org/abs/1902.10541}{{\ttfamily 1902.10541}}.

\bibitem{Delabrouille:2017rct}
{\scshape CORE} Collaboration, J.~Delabrouille et~al., \emph{{Exploring cosmic
  origins with CORE: Survey requirements and mission design}},
  \href{https://doi.org/10.1088/1475-7516/2018/04/014}{\emph{JCAP} {\bfseries
  04} (2018) 014} [\href{https://arxiv.org/abs/1706.04516}{{\ttfamily
  1706.04516}}].

\bibitem{Gouttenoire:2019kij}
Y.~Gouttenoire, G.~Servant and P.~Simakachorn, \emph{{Beyond the Standard
  Models with Cosmic Strings}},
  \href{https://doi.org/10.1088/1475-7516/2020/07/032}{\emph{JCAP} {\bfseries
  07} (2020) 032} [\href{https://arxiv.org/abs/1912.02569}{{\ttfamily
  1912.02569}}].

\bibitem{Rogers:2020ltq}
K.~K. Rogers and H.~V. Peiris, \emph{{Strong bound on canonical ultra-light
  axion dark matter from the Lyman-alpha forest}},
  \href{https://doi.org/10.1103/PhysRevLett.126.071302}{\emph{Phys. Rev. Lett.}
  {\bfseries 126} (2021) 071302}
  [\href{https://arxiv.org/abs/2007.12705}{{\ttfamily 2007.12705}}].

\bibitem{Hiramatsu:2010yu}
T.~Hiramatsu, M.~Kawasaki, T.~Sekiguchi, M.~Yamaguchi and J.~Yokoyama,
  \emph{{Improved estimation of radiated axions from cosmological axionic
  strings}}, \href{https://doi.org/10.1103/PhysRevD.83.123531}{\emph{Phys. Rev.
  D} {\bfseries 83} (2011) 123531}
  [\href{https://arxiv.org/abs/1012.5502}{{\ttfamily 1012.5502}}].

\bibitem{Gorghetto:2020qws}
M.~Gorghetto, E.~Hardy and G.~Villadoro, \emph{{More Axions from Strings}},
  \href{https://arxiv.org/abs/2007.04990}{{\ttfamily 2007.04990}}.

\bibitem{Preskill:1982cy}
J.~Preskill, M.~B. Wise and F.~Wilczek, \emph{{Cosmology of the Invisible
  Axion}}, \href{https://doi.org/10.1016/0370-2693(83)90637-8}{\emph{Phys.
  Lett. B} {\bfseries 120} (1983) 127}.

\bibitem{Abbott:1982af}
L.~F. Abbott and P.~Sikivie, \emph{{A Cosmological Bound on the Invisible
  Axion}}, \href{https://doi.org/10.1016/0370-2693(83)90638-X}{\emph{Phys.
  Lett. B} {\bfseries 120} (1983) 133}.

\bibitem{Dine:1982ah}
M.~Dine and W.~Fischler, \emph{{The Not So Harmless Axion}},
  \href{https://doi.org/10.1016/0370-2693(83)90639-1}{\emph{Phys. Lett. B}
  {\bfseries 120} (1983) 137}.

\bibitem{Turner:1985si}
M.~S. Turner, \emph{{Cosmic and Local Mass Density of Invisible Axions}},
  \href{https://doi.org/10.1103/PhysRevD.33.889}{\emph{Phys. Rev. D} {\bfseries
  33} (1986) 889}.

\bibitem{Lyth:1991ub}
D.~H. Lyth, \emph{{Axions and inflation: Sitting in the vacuum}},
  \href{https://doi.org/10.1103/PhysRevD.45.3394}{\emph{Phys. Rev. D}
  {\bfseries 45} (1992) 3394}.

\bibitem{Bae:2008ue}
K.~J. Bae, J.-H. Huh and J.~E. Kim, \emph{{Update of axion CDM energy}},
  \href{https://doi.org/10.1088/1475-7516/2008/09/005}{\emph{JCAP} {\bfseries
  09} (2008) 005} [\href{https://arxiv.org/abs/0806.0497}{{\ttfamily
  0806.0497}}].

\bibitem{OHare:2021zrq}
C.~A.~J. O'Hare, G.~Pierobon, J.~Redondo and Y.~Y.~Y. Wong, \emph{{Simulations
  of axionlike particles in the postinflationary scenario}},
  \href{https://doi.org/10.1103/PhysRevD.105.055025}{\emph{Phys. Rev. D}
  {\bfseries 105} (2022) 055025}
  [\href{https://arxiv.org/abs/2112.05117}{{\ttfamily 2112.05117}}].

\bibitem{Davis:1989nj}
R.~L. Davis and E.~P.~S. Shellard, \emph{{Do Axions Need Inflation?}},
  \href{https://doi.org/10.1016/0550-3213(89)90187-9}{\emph{Nucl. Phys. B}
  {\bfseries 324} (1989) 167}.

\bibitem{Yamaguchi:1998gx}
M.~Yamaguchi, M.~Kawasaki and J.~Yokoyama, \emph{{Evolution of axionic strings
  and spectrum of axions radiated from them}},
  \href{https://doi.org/10.1103/PhysRevLett.82.4578}{\emph{Phys. Rev. Lett.}
  {\bfseries 82} (1999) 4578}
  [\href{https://arxiv.org/abs/hep-ph/9811311}{{\ttfamily hep-ph/9811311}}].

\bibitem{Kawasaki:2014sqa}
M.~Kawasaki, K.~Saikawa and T.~Sekiguchi, \emph{{Axion dark matter from
  topological defects}},
  \href{https://doi.org/10.1103/PhysRevD.91.065014}{\emph{Phys. Rev. D}
  {\bfseries 91} (2015) 065014}
  [\href{https://arxiv.org/abs/1412.0789}{{\ttfamily 1412.0789}}].

\bibitem{Klaer:2017ond}
V.~B.~. Klaer and G.~D. Moore, \emph{{The dark-matter axion mass}},
  \href{https://doi.org/10.1088/1475-7516/2017/11/049}{\emph{JCAP} {\bfseries
  11} (2017) 049} [\href{https://arxiv.org/abs/1708.07521}{{\ttfamily
  1708.07521}}].

\bibitem{Buschmann:2021sdq}
M.~Buschmann, J.~W. Foster, A.~Hook, A.~Peterson, D.~E. Willcox, W.~Zhang and
  B.~R. Safdi, \emph{{Dark matter from axion strings with adaptive mesh
  refinement}}, \href{https://doi.org/10.1038/s41467-022-28669-y}{\emph{Nature
  Commun.} {\bfseries 13} (2022) 1049}
  [\href{https://arxiv.org/abs/2108.05368}{{\ttfamily 2108.05368}}].

\bibitem{Das:2006ht}
S.~Das and N.~Weiner, \emph{{Late Forming Dark Matter in Theories of Neutrino
  Dark Energy}}, \href{https://doi.org/10.1103/PhysRevD.84.123511}{\emph{Phys.
  Rev. D} {\bfseries 84} (2011) 123511}
  [\href{https://arxiv.org/abs/astro-ph/0611353}{{\ttfamily
  astro-ph/0611353}}].

\bibitem{Sarkar:2014bca}
A.~Sarkar, S.~Das and S.~K. Sethi, \emph{{How Late can the Dark Matter form in
  our universe?}},
  \href{https://doi.org/10.1088/1475-7516/2015/03/004}{\emph{JCAP} {\bfseries
  03} (2015) 004} [\href{https://arxiv.org/abs/1410.7129}{{\ttfamily
  1410.7129}}].

\bibitem{Das:2020nwc}
S.~Das and E.~O. Nadler, \emph{{Constraints on the epoch of dark matter
  formation from Milky Way satellites}},
  \href{https://doi.org/10.1103/PhysRevD.103.043517}{\emph{Phys. Rev. D}
  {\bfseries 103} (2021) 043517}
  [\href{https://arxiv.org/abs/2010.01137}{{\ttfamily 2010.01137}}].

\bibitem{Diamanti:2017xfo}
R.~Diamanti, S.~Ando, S.~Gariazzo, O.~Mena and C.~Weniger, \emph{{Cold dark
  matter plus not-so-clumpy dark relics}},
  \href{https://doi.org/10.1088/1475-7516/2017/06/008}{\emph{JCAP} {\bfseries
  06} (2017) 008} [\href{https://arxiv.org/abs/1701.03128}{{\ttfamily
  1701.03128}}].

\bibitem{Mehta:2020kwu}
V.~M. Mehta, M.~Demirtas, C.~Long, D.~J.~E. Marsh, L.~Mcallister and M.~J.
  Stott, \emph{{Superradiance Exclusions in the Landscape of Type IIB String
  Theory}},  \href{https://arxiv.org/abs/2011.08693}{{\ttfamily 2011.08693}}.

\bibitem{Unal:2020jiy}
C.~\"Unal, F.~Pacucci and A.~Loeb, \emph{{Properties of ultralight bosons from
  heavy quasar spins via superradiance}},
  \href{https://doi.org/10.1088/1475-7516/2021/05/007}{\emph{JCAP} {\bfseries
  05} (2021) 007} [\href{https://arxiv.org/abs/2012.12790}{{\ttfamily
  2012.12790}}].

\bibitem{Baryakhtar:2020gao}
M.~Baryakhtar, M.~Galanis, R.~Lasenby and O.~Simon, \emph{{Black hole
  superradiance of self-interacting scalar fields}},
  \href{https://arxiv.org/abs/2011.11646}{{\ttfamily 2011.11646}}.

\bibitem{Sikivie:2020zpn}
P.~Sikivie, \emph{{Invisible Axion Search Methods}},
  \href{https://doi.org/10.1103/RevModPhys.93.015004}{\emph{Rev. Mod. Phys.}
  {\bfseries 93} (2021) 015004}
  [\href{https://arxiv.org/abs/2003.02206}{{\ttfamily 2003.02206}}].

\bibitem{Irastorza:2018dyq}
I.~G. Irastorza and J.~Redondo, \emph{{New experimental approaches in the
  search for axion-like particles}},
  \href{https://doi.org/10.1016/j.ppnp.2018.05.003}{\emph{Prog. Part. Nucl.
  Phys.} {\bfseries 102} (2018) 89}
  [\href{https://arxiv.org/abs/1801.08127}{{\ttfamily 1801.08127}}].

\bibitem{OHare:2020wah}
C.~A.~J. O'Hare and E.~Vitagliano, \emph{{Cornering the axion with CP-violating
  interactions}},
  \href{https://doi.org/10.1103/PhysRevD.102.115026}{\emph{Phys. Rev. D}
  {\bfseries 102} (2020) 115026}
  [\href{https://arxiv.org/abs/2010.03889}{{\ttfamily 2010.03889}}].

\bibitem{Kaneta:2016wvf}
K.~Kaneta, H.-S. Lee and S.~Yun, \emph{{Portal Connecting Dark Photons and
  Axions}}, \href{https://doi.org/10.1103/PhysRevLett.118.101802}{\emph{Phys.
  Rev. Lett.} {\bfseries 118} (2017) 101802}
  [\href{https://arxiv.org/abs/1611.01466}{{\ttfamily 1611.01466}}].

\bibitem{Kalashev:2018bra}
O.~E. Kalashev, A.~Kusenko and E.~Vitagliano, \emph{{Cosmic infrared background
  excess from axionlike particles and implications for multimessenger
  observations of blazars}},
  \href{https://doi.org/10.1103/PhysRevD.99.023002}{\emph{Phys. Rev. D}
  {\bfseries 99} (2019) 023002}
  [\href{https://arxiv.org/abs/1808.05613}{{\ttfamily 1808.05613}}].

\bibitem{Arias:2020tzl}
P.~Arias, A.~Arza, J.~Jaeckel and D.~Vargas-Arancibia, \emph{{Hidden Photon
  Dark Matter Interacting via Axion-like Particles}},
  \href{https://arxiv.org/abs/2007.12585}{{\ttfamily 2007.12585}}.

\bibitem{Caputo:2022npg}
A.~Caputo, H.~Liu, S.~Mishra-Sharma, M.~Pospelov and J.~T. Ruderman, \emph{{A
  Stimulating Explanation of the Extragalactic Radio Background}},
  \href{https://arxiv.org/abs/2206.07713}{{\ttfamily 2206.07713}}.

\bibitem{Carr:2020pbh}
B.~Carr, K.~Kohri, Y.~Sendouda and J.~Yokoyama, \emph{{Constraints on
  Primordial Black Holes}},  \href{https://arxiv.org/abs/2002.12778}{{\ttfamily
  2002.12778}}.

\bibitem{Widrow:1989fe}
L.~M. Widrow, \emph{{The Collapse of Nearly Spherical Domain Walls}},
  \href{https://doi.org/10.1103/PhysRevD.39.3576}{\emph{Phys. Rev. D}
  {\bfseries 39} (1989) 3576}.

\bibitem{Deng:2017uwc}
H.~Deng and A.~Vilenkin, \emph{{Primordial black hole formation by vacuum
  bubbles}}, \href{https://doi.org/10.1088/1475-7516/2017/12/044}{\emph{JCAP}
  {\bfseries 12} (2017) 044}
  [\href{https://arxiv.org/abs/1710.02865}{{\ttfamily 1710.02865}}].

\bibitem{Haiman:2000ky}
Z.~Haiman and A.~Loeb, \emph{{What is the highest plausible redshift of
  luminous quasars?}}, \href{https://doi.org/10.1086/320586}{\emph{Astrophys.
  J.} {\bfseries 552} (2001) 459}
  [\href{https://arxiv.org/abs/astro-ph/0011529}{{\ttfamily
  astro-ph/0011529}}].

\bibitem{2021ApJ...907L...1W}
F.~{Wang}, J.~{Yang}, X.~{Fan}, J.~F. {Hennawi}, A.~J. {Barth}, E.~{Banados},
  F.~{Bian}, K.~{Boutsia}, T.~{Connor}, F.~B. {Davies}, R.~{Decarli}, A.-C.
  {Eilers}, E.~P. {Farina}, R.~{Green}, L.~{Jiang}, J.-T. {Li},
  C.~{Mazzucchelli}, R.~{Nanni}, J.-T. {Schindler}, B.~{Venemans}, F.~{Walter},
  X.-B. {Wu} and M.~{Yue}, \emph{{A Luminous Quasar at Redshift 7.642}},
  \href{https://doi.org/10.3847/2041-8213/abd8c6}{\emph{Astrophys.J.Lett.}
  {\bfseries 907} (2021) L1}
  [\href{https://arxiv.org/abs/2101.03179}{{\ttfamily 2101.03179}}].

\bibitem{Inayoshi:2019fun}
K.~Inayoshi, E.~Visbal and Z.~Haiman, \emph{{The Assembly of the First Massive
  Black Holes}},
  \href{https://doi.org/10.1146/annurev-astro-120419-014455}{\emph{Ann. Rev.
  Astron. Astrophys.} {\bfseries 58} (2020) 27}
  [\href{https://arxiv.org/abs/1911.05791}{{\ttfamily 1911.05791}}].

\bibitem{Latif}
M.~A. Latif, D.~J. Whalen, S.~Khochfar, N.~P. Herrington and T.~E. Woods,
  \emph{Turbulent cold flows gave birth to the first quasars},
  \href{https://doi.org/10.1038/s41586-022-04813-y}{\emph{Nature} {\bfseries
  607} (2022) 48}.

\bibitem{Davoudiasl:2021ijv}
H.~Davoudiasl, P.~B. Denton and J.~Gehrlein, \emph{{Supermassive Black Holes,
  Ultralight Dark Matter, and Gravitational Waves from a First Order Phase
  Transition}},
  \href{https://doi.org/10.1103/PhysRevLett.128.081101}{\emph{Phys. Rev. Lett.}
  {\bfseries 128} (2022) 081101}
  [\href{https://arxiv.org/abs/2109.01678}{{\ttfamily 2109.01678}}].

\bibitem{Balberg:2001qg}
S.~Balberg and S.~L. Shapiro, \emph{{Gravothermal collapse of selfinteracting
  dark matter halos and the origin of massive black holes}},
  \href{https://doi.org/10.1103/PhysRevLett.88.101301}{\emph{Phys. Rev. Lett.}
  {\bfseries 88} (2002) 101301}
  [\href{https://arxiv.org/abs/astro-ph/0111176}{{\ttfamily
  astro-ph/0111176}}].

\bibitem{Balberg:2002ue}
S.~Balberg, S.~L. Shapiro and S.~Inagaki, \emph{{Selfinteracting dark matter
  halos and the gravothermal catastrophe}},
  \href{https://doi.org/10.1086/339038}{\emph{Astrophys. J.} {\bfseries 568}
  (2002) 475} [\href{https://arxiv.org/abs/astro-ph/0110561}{{\ttfamily
  astro-ph/0110561}}].

\bibitem{Pollack:2014rja}
J.~Pollack, D.~N. Spergel and P.~J. Steinhardt, \emph{{Supermassive Black Holes
  from Ultra-Strongly Self-Interacting Dark Matter}},
  \href{https://doi.org/10.1088/0004-637X/804/2/131}{\emph{Astrophys. J.}
  {\bfseries 804} (2015) 131}
  [\href{https://arxiv.org/abs/1501.00017}{{\ttfamily 1501.00017}}].

\bibitem{Xiao:2021ftk}
H.~Xiao, X.~Shen, P.~F. Hopkins and K.~M. Zurek, \emph{{SMBH seeds from
  dissipative dark matter}},
  \href{https://doi.org/10.1088/1475-7516/2021/07/039}{\emph{JCAP} {\bfseries
  07} (2021) 039} [\href{https://arxiv.org/abs/2103.13407}{{\ttfamily
  2103.13407}}].

\bibitem{Choquette:2018lvq}
J.~Choquette, J.~M. Cline and J.~M. Cornell, \emph{{Early formation of
  supermassive black holes via dark matter self-interactions}},
  \href{https://doi.org/10.1088/1475-7516/2019/07/036}{\emph{JCAP} {\bfseries
  07} (2019) 036} [\href{https://arxiv.org/abs/1812.05088}{{\ttfamily
  1812.05088}}].

\bibitem{Duechting:2004dk}
N.~Duechting, \emph{{Supermassive black holes from primordial black hole
  seeds}}, \href{https://doi.org/10.1103/PhysRevD.70.064015}{\emph{Phys. Rev.
  D} {\bfseries 70} (2004) 064015}
  [\href{https://arxiv.org/abs/astro-ph/0406260}{{\ttfamily
  astro-ph/0406260}}].

\bibitem{Khlopov:2004sc}
M.~Y. Khlopov, S.~G. Rubin and A.~S. Sakharov, \emph{{Primordial structure of
  massive black hole clusters}},
  \href{https://doi.org/10.1016/j.astropartphys.2004.12.002}{\emph{Astropart.
  Phys.} {\bfseries 23} (2005) 265}
  [\href{https://arxiv.org/abs/astro-ph/0401532}{{\ttfamily
  astro-ph/0401532}}].

\bibitem{Kusenko:2020pcg}
A.~Kusenko, M.~Sasaki, S.~Sugiyama, M.~Takada, V.~Takhistov and E.~Vitagliano,
  \emph{{Exploring Primordial Black Holes from the Multiverse with Optical
  Telescopes}},
  \href{https://doi.org/10.1103/PhysRevLett.125.181304}{\emph{Phys. Rev. Lett.}
  {\bfseries 125} (2020) 181304}
  [\href{https://arxiv.org/abs/2001.09160}{{\ttfamily 2001.09160}}].

\bibitem{Harlow:2018tng}
D.~Harlow and H.~Ooguri, \emph{{Symmetries in quantum field theory and quantum
  gravity}}, \href{https://doi.org/10.1007/s00220-021-04040-y}{\emph{Commun.
  Math. Phys.} {\bfseries 383} (2021) 1669}
  [\href{https://arxiv.org/abs/1810.05338}{{\ttfamily 1810.05338}}].

\bibitem{Georgi:1981pu}
H.~M. Georgi, L.~J. Hall and M.~B. Wise, \emph{{Grand Unified Models With an
  Automatic {Peccei-Quinn} Symmetry}},
  \href{https://doi.org/10.1016/0550-3213(81)90433-8}{\emph{Nucl. Phys. B}
  {\bfseries 192} (1981) 409}.

\bibitem{DiLuzio:2020wdo}
L.~Di~Luzio, M.~Giannotti, E.~Nardi and L.~Visinelli, \emph{{The landscape of
  QCD axion models}},
  \href{https://doi.org/10.1016/j.physrep.2020.06.002}{\emph{Phys. Rept.}
  {\bfseries 870} (2020) 1} [\href{https://arxiv.org/abs/2003.01100}{{\ttfamily
  2003.01100}}].

\bibitem{Barr:1992qq}
S.~M. Barr and D.~Seckel, \emph{{Planck scale corrections to axion models}},
  \href{https://doi.org/10.1103/PhysRevD.46.539}{\emph{Phys. Rev. D} {\bfseries
  46} (1992) 539}.

\bibitem{Kamionkowski:1992mf}
M.~Kamionkowski and J.~March-Russell, \emph{{Planck scale physics and the
  Peccei-Quinn mechanism}},
  \href{https://doi.org/10.1016/0370-2693(92)90492-M}{\emph{Phys. Lett. B}
  {\bfseries 282} (1992) 137}
  [\href{https://arxiv.org/abs/hep-th/9202003}{{\ttfamily hep-th/9202003}}].

\bibitem{Kallosh:1995hi}
R.~Kallosh, A.~D. Linde, D.~A. Linde and L.~Susskind, \emph{{Gravity and global
  symmetries}}, \href{https://doi.org/10.1103/PhysRevD.52.912}{\emph{Phys. Rev.
  D} {\bfseries 52} (1995) 912}
  [\href{https://arxiv.org/abs/hep-th/9502069}{{\ttfamily hep-th/9502069}}].

\bibitem{Hiramatsu:2011eswn}
T.~Hiramatsu, M.~Kawasaki and K.~Saikawa, \emph{{Evolution of String-Wall
  Networks and Axionic Domain Wall Problem}},
  \href{https://doi.org/10.1088/j.jcap.2011.08.030}{\emph{Jour. Cosm. Astr.
  Phys.} {\bfseries 2011} (2011) 030}
  [\href{https://arxiv.org/abs/1012.4558}{{\ttfamily 1012.4558}}].

\end{thebibliography}\endgroup
 
\end{document}